\documentclass[twocolumn]{aastex631}
\received{\today}
\usepackage{gensymb}
\usepackage{substitutefont}
\substitutefont{TS1}{aer}{cmr}

\usepackage{tabularx}
\usepackage{url}
\usepackage{graphicx}
\usepackage[T1]{fontenc}
\usepackage{ae,aecompl}
\usepackage{booktabs}
\usepackage{natbib}
\usepackage{multirow}
\usepackage{multirow}
\usepackage{appendix}
\usepackage{overpic}
\usepackage{blindtext}
\usepackage{longtable}
\usepackage{tabu}
\usepackage{lineno}

\usepackage{chngcntr}

\newcommand\es{{1ES 1927+654}}

\def\aap{A\&A} \def\apjl{ApJ}  
 \def\apjs{ApJS}

\begin{document}

%\title{Late-time radio brightening in the changing-look AGN 1ES~1927+654}
\title{Late-time radio brightening and emergence of a radio jet in the changing-look AGN 1ES~1927+654}
\author[0000-0002-7676-9962]{Eileen T. Meyer}
\affiliation{Department of Physics, University of Maryland Baltimore County, 1000 Hilltop Circle Baltimore, MD 21250, USA}

\author[0000-0003-2714-0487]{Sibasish Laha} 

\affiliation{Astrophysics Science Division, NASA Goddard Space Flight Center, Greenbelt, MD 20771, USA.}
\affiliation{Center for Space Science and Technology, University of Maryland Baltimore County, 1000 Hilltop Circle, Baltimore, MD 21250, USA.}
\affiliation{Center for Research and Exploration in Space Science and Technology, NASA/GSFC, Greenbelt, Maryland 20771, USA}

\author[0000-0003-4727-2209]{Onic I. Shuvo}
\affiliation{Department of Physics, University of Maryland Baltimore County, 1000 Hilltop Circle Baltimore, MD 21250, USA}

\author[0000-0003-1101-8436]{Agniva Roychowdhury}
\affiliation{Space Telescope Science Institute, 3700 San Martin Dr, Baltimore, MD 21218, USA}
\affiliation{Department of Physics, University of Maryland Baltimore County, 1000 Hilltop Circle Baltimore, MD 21250, USA}

\author[0000-0003-3189-9998]{David A.\ Green}
\affiliation{Astrophysics Group, Cavendish Laboratory, 19 J.J. Thomson Avenue, Cambridge CB3 0HE, UK}

\author[0000-0003-2705-4941]{Lauren Rhodes}
\affiliation{Astrophysics, The University of Oxford, Keble Road, Oxford OX1 3RH, UK}

\author[0000-0001-9725-5509]{Amelia M. Hankla}
\altaffiliation{Neil Gehrels Fellow}
\affiliation{
Joint Space-Science Institute, University of Maryland, College Park, MD 20742, USA}
\affiliation{Department of Astronomy, University of Maryland, College Park, MD, USA}

\author[0000-0001-7801-0362]{Alexander Philippov}
\affiliation{
Department of Physics, University of Maryland, College Park, MD 20742, USA}

\author[0000-0001-9475-5292]{Rostom Mbarek}
\altaffiliation{Neil Gehrels Fellow}
\affiliation{
Joint Space-Science Institute, University of Maryland, College Park, MD 20742, USA}
\affiliation{Department of Astronomy, University of Maryland, College Park, MD, USA}

\author[0000-0002-1615-179X]{Ari laor}
\affiliation{Department of Physics, Technion, Haifa 32000, Israel}

\author[0000-0003-0936-8488]{Mitchell C.~Begelman}
\affiliation{JILA, University of Colorado and National Institute of Standards and Technology, 440 UCB, Boulder, CO 80309-0440, USA.}

\author[0000-0003-2714-0487]{Dev R. Sadaula} 
\affiliation{Astrophysics Science Division, NASA Goddard Space Flight Center, Greenbelt, MD 20771, USA.}
\affiliation{Center for Space Science and Technology, University of Maryland Baltimore County, 1000 Hilltop Circle, Baltimore, MD 21250, USA.}
\affiliation{Center for Research and Exploration in Space Science and Technology, NASA/GSFC, Greenbelt, Maryland 20771, USA}

\author[0000-0003-4790-2653]{Ritesh Ghosh}
\affiliation{Center for Space Science and Technology, University of Maryland Baltimore County, 1000 Hilltop Circle, Baltimore, MD 21250, USA.}
\affiliation{Astrophysics Science Division, NASA Goddard Space Flight Center, Greenbelt, MD 20771, USA.}
\affiliation{Center for Research and Exploration in Space Science and Technology, NASA/GSFC, Greenbelt, Maryland 20771, USA}
\affiliation{MKHS, Murshidabad, West Bengal, 742401, India.}

\author[0000-0002-5182-6289]{Gabriele Bruni}
\affiliation{INAF -- Istituto di Astrofisica e Planetologia Spaziali, Via del Fosso del Cavaliere 100, Roma, 00133, Italy}

\author[0000-0003-0543-3617]{Francesca Panessa}
\affiliation{INAF -- Istituto di Astrofisica e Planetologia Spaziali, Via del Fosso del Cavaliere 100, Roma, 00133, Italy}

\author[0000-0002-1094-3147]{Matteo Guainazzi}
\affiliation{European Space Agency (ESA), European Space Research and Technology Centre (ESTEC), Keplerlaan 1, 2201 AZ Noordwijk, The Netherlands}

\author[0000-0001-9735-4873]{Ehud Behar}
\affiliation{Department of Physics, Technion, Haifa 32000, Israel}
\affiliation{Department of Physics \& Kavli Institute for Astrophysics and Space Research, Massachusetts Institute of Technology, Cambridge, MA 02139, USA}

\author[0000-0003-4127-0739]{Megan Masterson}
\affiliation{Department of Physics \& Kavli Institute for Astrophysics and Space Research, Massachusetts Institute of Technology, Cambridge, MA 02139, USA}

\author[0000-0001-9826-1759]{Haocheng Zhang}
\affiliation{Center for Space Science and Technology, University of Maryland Baltimore County, 1000 Hilltop Circle, Baltimore, MD 21250, USA.}
\affiliation{Astrophysics Science Division, NASA Goddard Space Flight Center, Greenbelt, MD 20771, USA.}

\author[0000-0002-4439-5580]{Xiaolong Yang} 
\affiliation{Shanghai Astronomical Observatory, CAS, 80 Nandan Road, Shanghai 200030, China}

\author[0000-0003-0685-3621]{Mark A. Gurwell}
\affiliation{Center for Astrophysics | Harvard \& Smithsonian, 60 Garden Street, Cambridge, MA 02138, USA}

\author[0000-0002-3490-146X]{Garrett K. Keating}
\affiliation{Center for Astrophysics | Harvard \& Smithsonian, 60 Garden Street, Cambridge, MA 02138 USA}

\author[0000-0001-7361-0246]{David Williams-Baldwin}
\affiliation{Jodrell Bank Centre for Astrophysics, Department of Physics and Astronomy, The University of Manchester, Manchester M13 9PL, UK}

\author[0000-0002-0963-0223]{Justin D.\ Bray}
\affiliation{Jodrell Bank Centre for Astrophysics, Department of Physics and Astronomy, The University of Manchester, Manchester M13 9PL, UK}

\author[0000-0002-1727-1224]{Emmanuel K. Bempong-Manful}
\affiliation{Jodrell Bank Centre for Astrophysics, Department of Physics and Astronomy, The University of Manchester, Manchester M13 9PL, UK}
\affiliation{School of Physics, University of Bristol, Tyndall Avenue, Bristol BS8 1TL, UK}

\author{Nicholas Wrigley}
\affiliation{Jodrell Bank Centre for Astrophysics, Department of Physics and Astronomy, The University of Manchester, Manchester M13 9PL, UK}
   
\author[0000-0002-4622-4240]{Stefano Bianchi}
\affiliation{Dipartimento di Matematica e Fisica, Università degli Studi Roma Tre, via della Vasca Navale 84, I-00146 Roma, Italy}

\author[0000-0001-5742-5980]{Federica Ricci}
\affiliation{Dipartimento di Matematica e Fisica, Università degli Studi Roma Tre, via della Vasca Navale 84, I-00146 Roma, Italy}
\affiliation{INAF-Osservatorio Astronomico di Roma, via Frascati 33, 00040 Monteporzio Catone, Italy}

\author[0000-0002-1239-2721]{Fabio La Franca}
\affiliation{Dipartimento di Matematica e Fisica, Università degli Studi Roma Tre, via della Vasca Navale 84, I-00146 Roma, Italy}

\author[0000-0003-0172-0854]{Erin Kara}
\affiliation{Department of Physics \& Kavli Institute for Astrophysics and Space Research, Massachusetts Institute of Technology, Cambridge, MA 02139, USA}

\author[0000-0002-2040-8666]{Markos Georganopoulos}
\affiliation{Department of Physics, University of Maryland Baltimore County, 1000 Hilltop Circle Baltimore, MD 21250, USA}

\author[0000-0000-0000-0000]{Samantha Oates}
\affiliation{Birmingham Institute for Gravitational Wave Astronomy and School of Physics and Astronomy, University of Birmingham, Birmingham B15 2TT, UK}

\author[0000-0002-2555-3192]{Matt Nicholl}
\affiliation{Astrophysics Research Centre, School of Mathematics and Physics, Queens University Belfast, Belfast BT7 1NN, UK
}

\author[0000-0001-6523-6522]{Main Pal}
\affiliation{Department of Physics, Sri Venkateswara College, University of Delhi, Benito Juarez Road, Dhaula Kuan,  New Delhi -- 110021, India}

\author[0000-0003-1673-970X]{S. Bradley Cenko}
\affiliation{Astrophysics Science Division, NASA Goddard Space Flight Center, Greenbelt, MD 20771, USA.}
\affiliation{Joint Space-Science Institute, University of Maryland, College Park, MD 20742, USA}

\begin{abstract}
We present multi-frequency (5--345 GHz) and multi-resolution radio observations of 1ES~1927+654, widely considered one of the most unusual and extreme changing-look active galactic nuclei (CL-AGN). The source was first designated a CL-AGN after an optical outburst in late 2017 and has since displayed considerable changes in X-ray emission, including the destruction and rebuilding of the X-ray corona in 2019--2020. Radio observations prior to 2023 show a faint and compact radio source  typical of radio-quiet AGN.
Starting in February 2023, 1ES~1927+654 began exhibiting a radio flare with a steep exponential rise, reaching a peak 60~times previous flux levels, and has maintained this higher level of radio emission for  over a year to date. The 5--23 GHz spectrum is broadly similar to gigahertz-peaked radio sources, which are understood to be young radio jets less than $\sim$1000 years old.  Recent high-resolution VLBA observations at 23.5 GHz now show resolved extensions on either side of the core, with a separation of $\sim$0.15 pc, consistent with a new and mildly relativistic bipolar outflow.  A steady increase in the soft X-ray band (0.3--2 keV) concurrent with the radio may be consistent with jet-driven shocked gas, though further observations are needed to test alternate scenarios. This source joins a growing number of CL-AGN and tidal disruption events which show late-time radio activity, years after the initial outburst.

\end{abstract}

\keywords{Radio AGN -- X-ray AGN -- Seyfert galaxies -- jets -- proper motion}

\vspace{50pt}
\section{Introduction} 
\label{sec:intro}

Recent advances in time-domain studies have led to the identification of new types of extreme variability in active galaxies, popularly called ``changing look'' active galactic nuclei (CL-AGNs, hereafter). These extreme variations are not only characterized by orders-of-magnitude changes in the optical, UV, and X-ray luminosity of the source but also by an unexpected transition between optical spectral type~\citep{Mathur_2018ApJ...866..123M,Trakhtenbrot_2019ApJ...883...94T,Kokubo_2020MNRAS.491.4615K,Komossa_2020A&A...643L...7K}. In the simplest AGN unification framework, those with broad emission lines (type I) are thought to be viewed more face-on, such that the central nucleus is unobscured, while type II AGN are thought to be viewed at higher inclination angles, in which cases an inferred dusty torus obscures the broad-line emitting clouds, resulting in only narrow optical emission lines in the optical spectrum \citep{antonucci1993,bianchi2012,ram2017}. While we have long known that many factors complicate this simple picture (e.g. strength of the central continuum, variations in the amount and distribution of molecular gas and dust), the drastic changes seen in CL-AGNs in such short timescales (a few weeks to months) challenge the simplest form of the AGN unification framework which has generally assumed that such changes occur over far longer timescales \citep[see][for a recent review]{ricci2023}.

The AGN 1ES~1927+654 ($z=0.017$; 348\,pc/$''$) is widely considered one of the most unusual and extreme CL-AGN yet discovered \citep{Trakhtenbrot_2019ApJ...883...94T, ricci2020, ricci2021,masterson2022, Laha_2022ApJ...931....5L, Ghosh_2023ApJ...955....3G}, and is one of the extremely few CL-AGN observed to change in real time.  Earlier X-ray and optical observations of \es\ classified it as a ``true'' or ``naked'' type II AGN, with no evidence for obscuration by dust along the line of sight to explain the lack of broad emission lines \citep{boller2003,tran2011}. It was first flagged as a transient source of significantly increasing flux by the All-Sky Automated Survey for Supernovae (ASAS-SN;~\citealp[]{Shappee_2014ApJ...788...48S}) program on 2018 March~3 \citep[ASAS-SN-18el/AT2018zf;][]{nicholls2018}. In the original discovery paper, \citet{Trakhtenbrot_2019ApJ...883...94T} used archival data from the Asteroid Terrestrial-impact
Last Alert System (ATLAS;~\citealp[]{Trony_2018PASP..130f4505T}) to show that the outburst actually began in December 2017 and that the total increase was nearly 5~magnitudes in the V~band (a factor of 100 in total flux). 
An initially nearly featureless optical quasar spectrum began to show increasing broad lines approximately 70$-$90 days after the peak in the optical band, which occurred in March 2018. The broad line fluxes continued to increase in strength out to roughly 150 days after the continuum peak, and persisted
for at least 11~months after the optical flare.  This behavior follows the expectations for the illumination of a previously existing broad line region of size $\sim$1$-$3 light-months.

Subsequent to the initial CL outburst, the source has displayed a series of unusually varied states in the X-ray band while showing no change other than a monotonic return to pre-CL values in the optical/UV and general quiescence in the radio (up to the presently reported outburst). After a short period of X-ray emission similar to pre-CL levels, the 2--10 keV hard X-ray emission (i.e.\ the corona) completely vanished for about 3~months in 2018 before flaring up by a factor 1000 to exceed the Eddington limit for a $10^6$ $M_\odot$ black hole \citep{ricci2020}, a state it maintained for over a year before dropping back to pre-outburst levels \citep{ricci2021, masterson2022}. 

While the inital optical/UV flare and decay timescale appear consistent with a possible tidal disruption event (TDE) in an existing AGN, the X-ray spectral changes are not, including both short and long-term variability and temperature variations for the thermal X-ray component in the first few years after the CL event
\citep{masterson2022}. One possibility is that the CL event was the result of a magnetic flux inversion event in a magnetically arrested disk or MAD \citep{scepi2021}. This could explain the much shallower UV flux decay than expected under a TDE, the X-ray minimum following the outburst, and the lack of TDE-like spectral features  \citep{Laha_2022ApJ...931....5L}.  Most recently, \cite{Ghosh_2023ApJ...955....3G} have reported the emergence of a new bright and soft X-ray component in 1ES~1927+654 (along with an undiminished X-ray corona), which began in late 2022 and continues up to the present epoch. It was this enhanced state which triggered a director's discretionary time (DDT) request to the Very Long Baseline Array (VLBA) which led to the fortunate timing of much of the radio monitoring presented here.

\section{Observations} \label{sec:Methodology}

\subsection{Overview of Past and Present Radio Monitoring}

The source 1ES~1927+654 has been observed in the radio with the Very Large Array (VLA) and Very Long Baseline Array (VLBA) on only a few occasions prior to the CL event. \citet{Perlman_1996ApJS..104..251P} report a flux density of 16 mJy at 6 cm measured by the VLA in CnB hybrid configuration with a resolution of approximately 5$''$. Observations with the VLBA taken in 2013 and 2014 \citep[previously published in][]{Laha_2022ApJ...931....5L} are included here for comparison to more recent data. Post-CL event, the source was observed by the VLBA once at C-band in December 2018 and then sporadically with both the VLBA and European VLBI Network (EVN) on 6--12 month timescales from January 2020 by our group and others. This longer-timescale monitoring was primarily motivated by the months-timescale X-ray changes seen by \emph{Swift} and \emph{NICER}, to look for possible correlated radio variability. 

While our standard monitoring observations with the VLBA at 5 GHz showed no major change between 2022 March and August (total flux of 6 and 5 mJy, respectively), we made a director's discretionary time (DDT) request in April 2023 (BR 256) based on the soft X-ray flux seen by \emph{Swift}, which had been steadily rising since late 2022. This observation resulted in the surprising finding that the C-band (5 GHz) flux density had increased by a factor of 7 to 37.4 mJy. We immediately triggered additional DDT requests to the VLBA, EVN, VLA, AMI, e-MERLIN, and the SMA from April 2023 up to the time of this paper, on timescales ranging from every few days to monthly. As of May 2024, the source is still being monitored regularly by the VLBA, EVN, and e-MERLIN. In this initial publication, we present primarily the VLBA and EVN observations of 1ES~1927+654 to date, as well as the observations by AMI, one of the several VLA observations obtained in May 2023, and  mm-band measurements by the SMA in July 2023 and June 2024. 

\begin{figure}[t]
    \centering
    \includegraphics[width=0.48\textwidth]{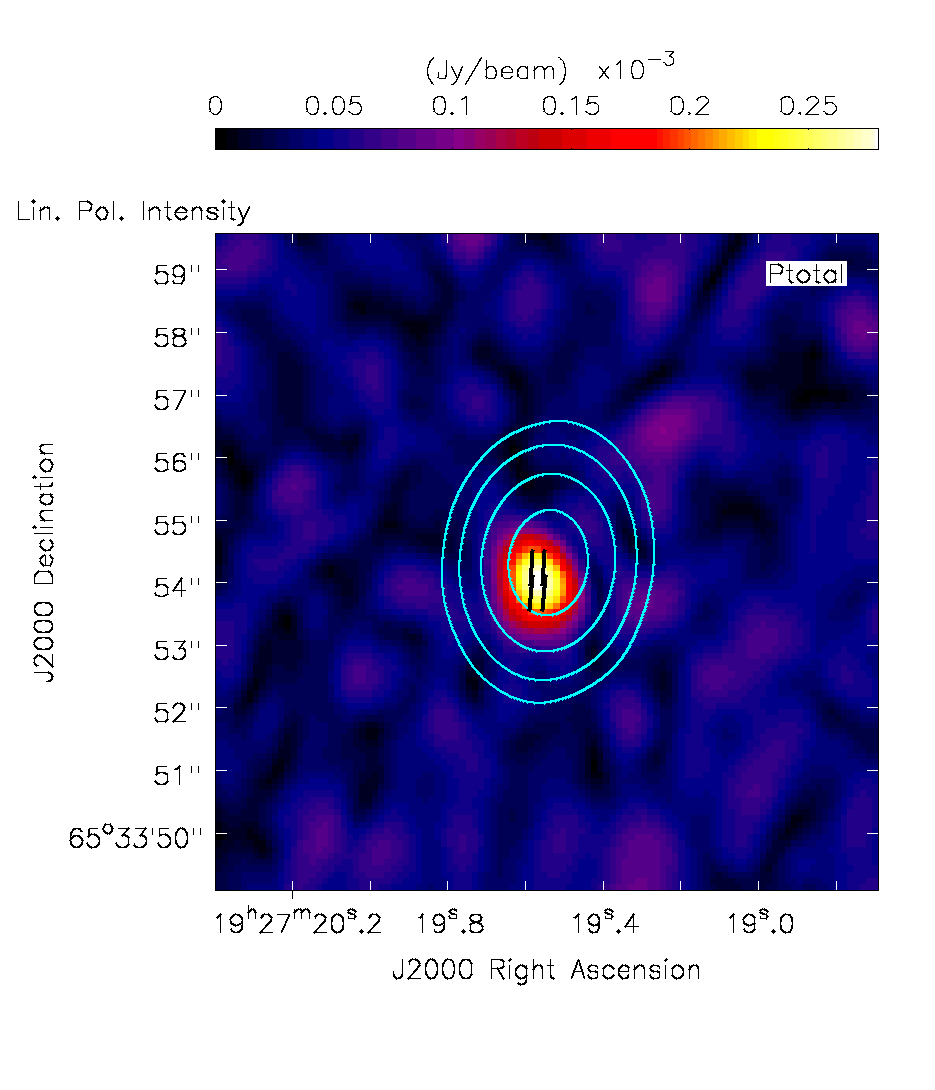}
    \caption{Total polarized intensity map at 5.5 GHz for the VLA observation of 2023-05-21. Color scale gives the flux density in mJy\,beam$^{-1}$. Contours overlaid in cyan correspond to the total (Stokes~I) intensity, with levels at 10, 50, 250 and 1000 times the base value 2.5$\times10^{-5}$ Jy\,beam$^{-1}$ which is approximately the total intensity image RMS. The vectors shown in black correspond to the measured linear polarization EVPA (no rotation applied). The polarized intensity peak appears slightly offset from the total intensity peak and has a value of 0.29 mJy, corresponding to a 0.6\% polarization fraction.}
    \label{fig:vla}
\end{figure}

\subsection{VLBA Calibration and Imaging}  \label{sec:vlba_cal}

The VLBI observations spanning 2013 to early 2024 are summarized in Table~\ref{tab:Observation}, where we describe both the VLBA observations and those from the European VLBI Network (EVN), discussed in the next section. In particular we give the observation date, program name, central frequency in GHz, and the final image RMS value in mJy/beam in columns 1-5. In columns 6 and 7 we give the peak radio flux in mJy/beam and the total radio flux of the source in mJy. The restoring beam (i.e., approximately the resolution of the imaging) is given in column 8 as the semi-major and minor ellipse axis length in mas, and the orientation of the ellipse in degrees in column 9. The final column notes whether self-calibration was possible for the imaging (`no' indicates it was not, while `p' indicates phase-only self-calibration was applied, and `ap' indicates both amplitude and phase).

\clearpage

\startlongtable
\begin{deluxetable*}{llcrcccccc}
%\tablenum{2}   
\small
\label{tab:Observation}
\tablecaption{VLBA and EVN Observation and Radio Properties.}

\tablehead{\colhead{Date} &  \colhead{Segment} &  \colhead{Band} & \colhead{Freq.} & \colhead{RMS}   & \colhead{$F_\mathrm{peak}$} & \colhead{$S_\mathrm{tot}$} & \colhead{Rest.\ Beam} & \colhead{Beam Angle} & Self-calibration\\[-1ex]
        \colhead{}       & \colhead{}     &  \colhead{}	   & \colhead{(GHz)} & \colhead{(mJy bm$^{-1}$)} & \colhead{(mJy\,bm$^{-1}$)} & \colhead{(mJy)}    &        \colhead{($\alpha \times \delta$; mas)}  &  \colhead{(deg)} & \\[-1ex]
        %\colnumbers}
        }
        \colnumbers
\startdata
\multicolumn{8}{c}{\emph{--Previously Published}--} \\
2013-08-10 & EG079A$^{\dagger}$ & L   & 1.48   & 0.25  & \nodata  & 18.9& 28.2$\times$11.7 & \nodata& \nodata\\
2014-03-25 & EG079B$^{\dagger}$   & C   & 4.99   &  0.03 & 3.6$^*$ & 4.1 & 2.47$\times$1.18 & \nodata& \nodata\\
2018-12-04     & RSY07$^{\dagger}$ & C   & 4.99   & 0.04  & 0.7$^*$ & 8.4 & 6.01$\times$4.95 & \nodata& \nodata\\
\hline
\multicolumn{8}{c}{\emph{--New Observations}--} \\
2020-01-17  & BY149     & S  & 2.28 &0.15	&1.78 &3.99   & 6.4$\times$3.3 & \phn~ 0.95&no\\[-0.8ex]
            &            & X  & 8.43 &0.03	&0.26 &1.61   & 1.7$\times$1.0 & \phn~ 5.46&no\\
2021-03-15  & BM518     & C  & 4.98 &0.04	&1.33 &5.32   & 3.9$\times$1.5 & \phn 15.82 &no\\
2022-03-05  & BM527     & C  & 4.87 &0.04	&2.95 &5.89   & 3.8$\times$2.5 & \phn 28.98&no\\[-0.8ex]
            &            & X  & 8.37 &0.03	&1.20 &2.38   & 2.2$\times$1.2 & \phn 22.59 &no\\
2022-08-05  & BY177B    & C  & 4.87 &0.03	&1.89 &4.95   & 4.0$\times$1.6 & \phn17.88 &no\\
2022-08-08  & BY177C    & X  & 8.37 &0.02	&1.10 &1.83   & 1.6$\times$0.9 & \phn$-$4.23&no\\
2023-02-10  & BY177F    & X  & 8.37 &0.03	&10.57 &11.92  & 1.7$\times$1.0 &  $-$11.48&p\\
2023-02-18  & BY177E    & C  & 4.87 &0.21	&6.84 &8.26   & 4.3$\times$2.0 & \phn13.46 &p\\
2023-04-27  & BR256     & C  & 4.98 &0.05	&36.32 &37.47  & 3.9$\times$1.5 & \phn$-$5.15&a$+$p\\[-0.8ex]
            &            & X  & 8.42 &0.05	&47.09 &48.77  & 2.3$\times$0.9 & \phn$-$8.06& a$+$p\\
2023-05-21  & BM549A    & C  & 4.98 &0.07	&48.44 &49.74  & 3.9$\times$1.4 & \phn$-$4.15&a$+$p\\[-0.8ex]
            &            & X  & 8.42 &0.04	&39.79 &40.61  & 2.3$\times$0.9 & \phn$-$6.76 &a$+$p\\
2023-05-28  & BM549B    & C  & 4.98 &0.05	&53.33 &55.92  & 4.1$\times$1.5 & \phn$-$4.15 &a$+$p \\[-0.8ex]
            &            & X  & 8.42 &0.07	&45.91 &47.32  & 2.5$\times$1.0 & \phn$-$5.80&a$+$p\\
2023-06-01  &BM549C     & C  & 4.98 &0.05	&67.36 &67.60  & 3.9$\times$1.6 & \phn$-$9.18 &a$+$p\\[-0.8ex]
            &            & X  & 8.42 &0.05	&50.11 &51.76  & 2.3$\times$0.8 & \phn$-$9.94 &a$+$p\\
2023-06-08  &BM549D     & C  & 4.98 &0.10	&69.27 &73.10  & 4.0$\times$1.8 & \phn~20.75 &a$+$p\\[-0.8ex]
            &            & X  & 8.42 &0.13	&57.22 &58.58  & 2.2$\times$0.8 & \phn~23.17 &a$+$p\\[-0.8ex]
            &            & K  & 22.22&0.11	&20.69 &20.91  & 0.9$\times$0.4 & \phn~19.07 &a$+$p\\
2023-06-28  &RB008$^\dagger$ & C &4.93 &0.08 & 55.59& 54.58& 9.2$\times$2.7  & $-$85.89&a$+$p\\
2023-07-22  &BM550A     & C  & 4.87 &0.07	&58.42 &63.82  & 4.0$\times$1.8 & $-$22.15 &a$+$p\\[-0.8ex]
            &            & X  & 8.37 &0.10	&59.42 &61.29  & 2.4$\times$1.0 & $-$25.01 &a$+$p\\[-0.8ex]
            &            & K  & 23.55&0.21	&10.90 &13.05   & 0.8$\times$0.4 & $-$31.35 &no\\
2023-08-31  &BM550B     & C  & 4.87 &0.12	&52.06 &54.31  & 3.6$\times$1.3 & $-$19.58&a$+$p\\[-0.8ex]
            &            & X  & 8.36 &0.07	&53.85 &55.65  & 2.3$\times$0.8 & $-$23.72&a$+$p\\[-0.8ex]
            &            & K  & 23.57&0.30	&5.66 &5.27   & 0.9$\times$0.3 & $-$27.93&no\\
2023-09-23  &BM550C     & C  & 4.87 &0.12	&62.26 &65.50  & 4.6$\times$1.6 & $-$15.75&a$+$p\\[-0.8ex]
            &            & X  & 8.36 &0.11	&52.07 &54.55  & 2.8$\times$0.9 & $-$17.57&a$+$p\\[-0.8ex]
            &            & K  & 23.57&0.30	&5.06 &6.30   & 1.0$\times$0.4 & $-$22.25&no\\
2023-10-22  &EY038A$^{\dagger}$      & X  & 8.41 &0.06 &44.53&47.51 & 1.5$\times$0.5&$-$44.84&a$+$p\\
2023-10-26  &EY038B$^{\dagger}$      & L  & 1.66 &0.05 &32.67 &45.81 & 15.9$\times$3.1&$-$47.47&a$+$p\\
2023-10-27  &BM550D     & C  & 4.87 &0.06	&64.73 &65.07  & 4.3$\times$2.0 & $-$26.28&a$+$p\\[-0.8ex]
            &            & X  & 8.36 &0.06	&50.88 &54.50  & 2.5$\times$1.0 & $-$25.48&a$+$p\\[-0.8ex]
            &            & K  & 23.57&0.24	&7.45 &19.50  & 0.9$\times$0.6 & \phn~27.04&no\\
2023-11-01  &EY038C$^{\dagger}$      & C  & 4.93&0.03 &48.20 &51.07 &4.2$\times$2.7 &\phn~32.38&a$+$p\\
2023-11-09  &RB009$^\dagger$ & K &22.24&0.50&9.16 & 10.45 & 0.7$\times$0.2 & \phn~ 9.22&no\\
2023-11-26  &BM550E     & C  & 4.87 &0.07	&54.60 &59.66  & 4.1$\times$1.9 & $-$20.92&a$+$p\\[-0.8ex]
            &            & X  & 8.37 &0.20	&45.31 &46.68  & 2.3$\times$0.8 & $-$21.00&a$+$p\\[-0.8ex]
            &            & K  & 23.57&0.07	&13.12 &15.86  & 0.9$\times$0.4 & $-$23.83 &a$+$p\\
2023-12-26  &BM550F     & C  & 4.87 &0.08	&64.82 &69.02  & 4.4$\times$1.5 & $-$21.44&a$+$p\\[-0.8ex]
            &            & X  & 8.37 &0.07	&48.99 &51.40  & 2.6$\times$0.8 & $-$24.06&a$+$p\\[-0.8ex]
            &            & K  & 23.57&0.12	&9.93 &14.77  & 0.9$\times$0.4 & $-$32.18 &p\\
2024-02-09  &BM556A     & C  & 4.87 &0.06	&69.64 &74.36  & 3.0$\times$2.0 & $-$33.80&a$+$p\\[-0.8ex]
            &            & X  & 8.37 &0.05	&47.42 &50.44  & 1.8$\times$1.1 & $-$31.59&a$+$p\\[-0.8ex]
            &            & K  & 23.57&0.11	&8.15 &10.53  & 0.7$\times$0.5 & $-$43.13 &p\\
2024-03-12  &EB106$^\dagger$ &K&22.24&0.16&6.16 & 7.09 & 1.0$\times$0.7 & \phn~$-$7.68&no\\
2024-04-24  &BM556B      & C  & 4.87 &0.08  &69.30&74.47  &4.6$\times$1.4 & \phn~$-$0.47&a$+$p\\[-0.8ex]
            &            & X  & 8.37 &0.05	&54.04 & 57.97  & 2.7$\times$0.9 & \phn~$-$6.84&a$+$p\\[-0.8ex]
            &            & K  & 23.57 &0.14  &17.10 & 19.52  &0.9$\times$0.4 & \phn~$-$11.61&p\\
2024-05-30  &BM556C      & L  & 1.63 &0.36  &29.60&36.50  &9.2$\times$4.3 & \phn~$-$28.70&a$+$p\\[-0.8ex]
            &            & C  & 4.87 &0.29  &74.72&77.23 &3.3$\times$1.6 & \phn~$-$20.75&a$+$p\\[-0.8ex]
            &            & X  & 8.37 &0.12	&61.36 & 66.42  & 1.9$\times$0.9 & \phn~$-$24.68&a$+$p\\[-0.8ex]
            &            & K  & 23.57 &0.06  &19.88 & 24.38  &0.8$\times$0.4 & \phn~$-$24.15& p\\
\enddata
\tablecomments{\footnotesize Previously published observations (top 3~entries) taken from \citet{Laha_2022ApJ...931....5L}. \\
$^{\dagger}$Observed by the European VLBI Network (EVN).\\
$^*$fluxes were reported as central point source (PS) flux densities. 
%$\ddagger$ 
}
\end{deluxetable*} 

For those observations with sufficient signal/noise ratio (SNR) per antenna for reasonable short solution intervals, we applied one or more iterations of phase-only self-calibration. In some cases the SNR was insufficient to proceed further, while in others (most C and X band observations) a single additional amplitude and phase self-calibration was applied. We verified that any self-calibration applied resulted in improved image RMS, and the values reported in Table 1 correspond to the self-calibrated image where this is the case.

In the initial calibration of all VLBA observations, we utilized the National Radio Astronomy Observatory (NRAO) Astronomical Image Processing System, also known as AIPS~\citep[][]{Van_1996ASPC..101...37V}. Specifically, we used the new primary development version of AIPS, release 31DEC23. Each frequency dataset was calibrated independently by pairing the target source with the phase calibrator (J1933+6540). We followed standard calibration procedures using \textsc{VLBAUTIL} and flagged bad data when necessary. We completed the calibration process for the phase calibrator and applied the calibrations to the source using the {\sc{split}} task. Once these standard AIPS calibration procedures were finished, we proceeded to the imaging stage. We utilized the {\sc imagr} task in AIPS to create images of both the calibrators and sources. 

We also utilized the NRAO Common Astronomical Software Applications \citep[\textsc{casa};][]{Casa_2022PASP..134k4501C}\footnote{\url{https://casa.nrao.edu}} to analyze images from our VLBA observations. To determine the integrated flux densities for sources, we used the two-dimensional fitting application through the {\sc viewer} command in \textsc{casa}. In Table~\ref{tab:Observation}, we listed the flux density values for all of our observations, along with all the radio properties of the final images.

In the three most recent K-band VLBA observations (9 Feb, 24 Apr and 30 May 2024), resolved components were apparent in the residual images after initial core was subtracted during the CLEAN deconvolution (further discussed in Section \ref{ext}). A more extensive investigation into the source structure at all bands (and complimented by the necessary observation simulations to provide better estimates of systematic errors) is planned for a future publication, but we do present some initial results here, focusing on the highest resolution K-band data.  In order to measure the flux and position of the resolved components, we produced core-subtracted measurement sets and images. Beginning in each case from the original data with one round of phase-only self-calibration applied,  we used a small clean box region of only a few pixel extent over the core region to produce a point-source model in the measurement set corresponding to the core emission only\footnote{From the variability timescale we know that the central source which is responsible for the fast rise is very small, well below the VLBA imaging resolution and thus appropriately modeled as a point source; typical point source localization accuracy is on the order of a few to tens of $\mu$as.}, stopping at the same residual flux level (approx.\ 0.19 mJy\,beam$^{-1}$) in the central core region. For comparison we also produced a core-subtracted image for the June 2023 observation, continuing to the level of the image RMS (since no residual emission  was apparent). The core emission was then subtracted using the CASA task \texttt{uvsub}, before re-imaging the resulting measurement set.  We utilized \texttt{DIFMAP} \citep{1997ASPC..125...77S} on the core-subtracted K band visibility data from the 2024 K-band epochs to test the presence of true extended emission. All of the epochs were consistent, to first order, with a dual point source structure located on opposite sides of the core location. The evolution of the flux densities and the positions of the point sources have been noted in Table 2, along with the separation between the two point sources.

\subsection{EVN Calibration and Imaging}  \label{sec:evn_cal}

The EVN epochs performed between 2023 and 2024 are summarized in Table~\ref{tab:Observation}. Phase-referencing was performed, adopting the same calibrator as for the 
VLBA observations (J1933+6540, at a separation of $\sim$0.7 deg). For the first epoch (RB008, C band) participating antennas were Jb,Wb,Ef,Mc,Nt,O8,T6,Tr,Hh,Ys and Ib, while the later epochs at higher frequency (RB009 and EB106, K band), could count on the participation of the Korean VLBI Network (KVN) adding sensitivity to the longest baselines. Participating antennas were Jb, Ef, Mc, O6, T6, Ur, Tr, Mh, Kt, Ky, Ku and Jb, Ef, Mc, O6, T6, Ur, Tr, Mh, Ys, Sr, Kt, Ky, Ku, respectively. Further information and observations logs are available on the EVN archive (https://archive.jive.nl/scripts/portal.php). Raw data from single antennas were correlated at the Joint Institute for VLBI ERIC (JIVE, Dwingeloo). Calibration was performed in {\tt{CASA}}, through the {\tt{rPICARD}} pipeline \citep{picard}. The calibrated visibilities were then imaged with {\tt{DIFMAP}} \citep{1997ASPC..125...77S}, following a standard phase and amplitude self-calibration procedure. Finally, integrated flux densities were extracted via a two-dimensional Gaussian fitting in {\tt{CASA}}.

Three additional EVN epochs—EY038A (observed on October 22, 2023), EY038B (October 26, 2023), and EY038C (November 1, 2023) were obtained from the PI (X. Yang) after our initial analysis of the above-referenced data. These observations followed similar phase-referencing techniques with the same calibrator J1933+6540. The antennas involved were as follows: EY038A included WB, EF, MC, O6, T6, UR, TR, IB, and WZ; EY038B included JB, WB, EF, MC, O8, T6, UR, TR, and IR; and EY038C included JB, WB, EF, MC, O8, T6, UR, TR, IB, and SR. Calibration followed the procedures outlined in the EVN data reduction guide  (https://www.evlbi.org/evn-data-reduction-guide), with imaging and self-calibration carried out similarly to the other EVN and VLBA observations.

\subsection{Very Large Array C-band Observations}

In the high-cadence VLA observing program 23A-407, our source was observed 11 times between May 20 and June~1. Here we present the observations of 21 May 2023 which were contemporaneous with a VLBA observation, to put limits on the large-scale radio flux which may contaminate lower-resolution observation (such as with AMI), and to measure the degree of polarization in the radio, if any. A full presentation of the remaining VLA observations will be made in a forthcoming paper. 

For the 21 May 2023 observations the array was in B configuration and we used a standard 8-bit C-band continuum observing setup with 2048 MHz bandwidth. Radio frequency interference (RFI) losses resulted in a reduction of bandwidth by an estimated 15--20\% on most baselines and the complete loss of spectral window~10. 3C~48 served as the primary bandpass and flux calibrator, as well as the polarization calibrator for the cross-hand delay and polarization angle, using values from 2019 available from the NRAO website \citep[see e.g.][]{perley2017}. The unpolarized source J2355+4950 was used to determine instrumental polarization (leakage/D-terms), and J1927+6117 was used as a secondary gain calibrator. All calibration, imaging, and analysis was carried out with the CASA package. We used a modified version of the CASA pipeline (6.5.4.9) for the initial calibration and then determined the polarization calibration tables by hand after additional flagging of both source and calibrators for RFI. In the initial pipeline run we ran the \texttt{hifv\_syspower} task with \texttt{apply=True} to counteract the effects of gain compression due to strong RFI.
The CASA task \texttt{tclean} was used in interactive mode for imaging deconvolution using standard wide-band continuum parameters (mtmfs deconvolver with nterms=2, natural weighting). Two rounds of self-calibration were applied after initial imaging of Stokes' I only, one phase-only followed by a cumulative amplitude+phase table. The final full-Stokes (IQUV) imaging used natural weighting with a pixel scale of $0\farcs1$. The resulting images have a restoring beam of $1\farcs6 \times 1\farcs2$ oriented at $-4.25^\circ$. The Stokes' I, Q, U, and V images have RMS values of 2.5, 2.1, 2.4, and $2.2 \times 10^{-5}$ Jy\,beam$^{-1}$, respectively. 

\begin{figure*}[t]
    \centering
    \includegraphics[width=0.95\textwidth]{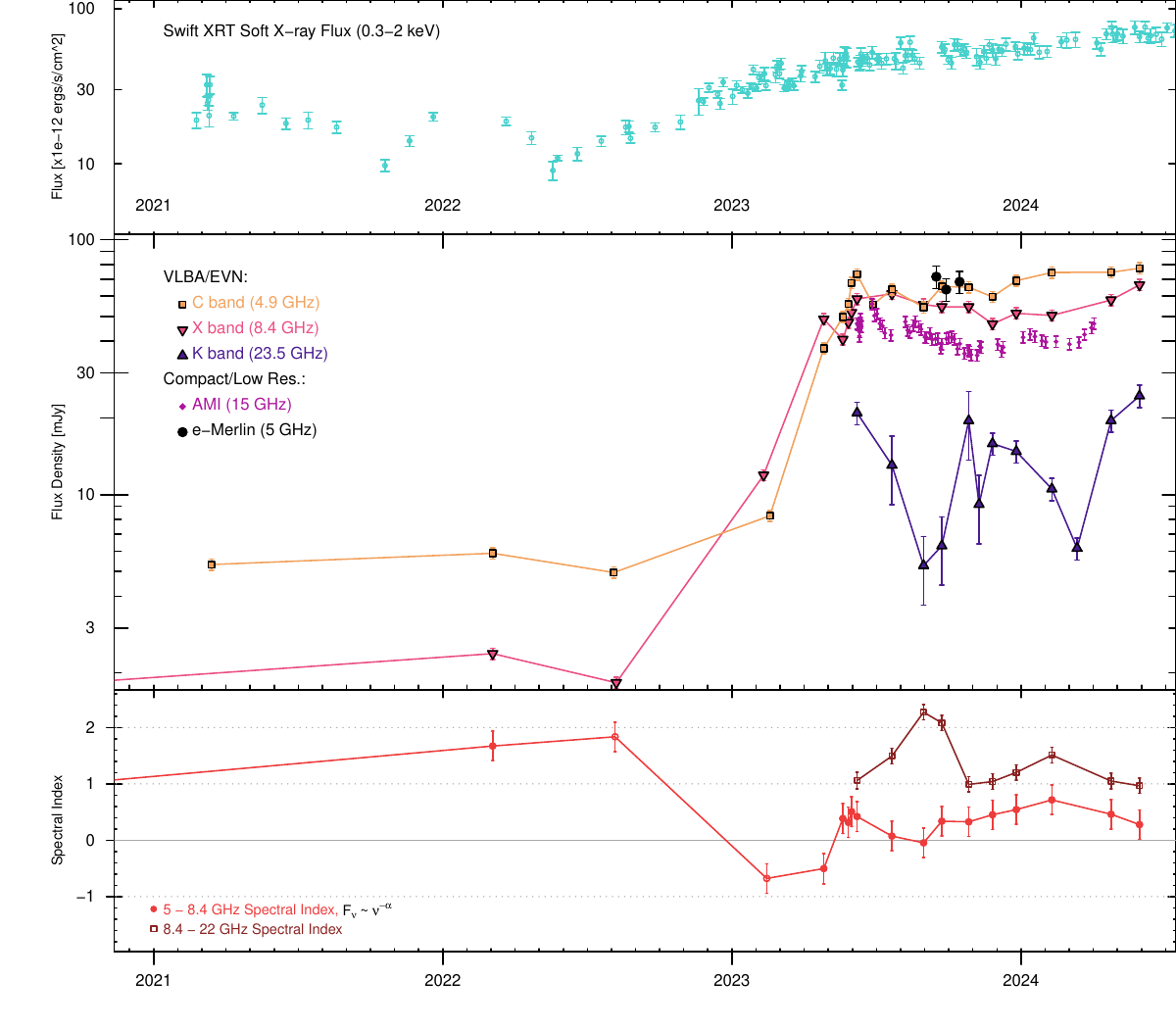}
    \caption{The soft X-ray and radio lightcurves of 1ES~1927+654 since 2021. The top panel shows the 0.3--2 keV flux (in units of $10^{-12}\rm erg\, cm^{-2}\, s^{-1}$) observed by the \emph{Swift}/XRT \citep[further details in][]{laha2024}; central panel shows the total (log scale) VLBA/EVN flux in bands C, X, and K (5, 8.4, and 22.2/23.5 GHz, respectively) along with fluxes from lower-resolution AMI and e-MERLIN observations at 15.5 and 5 GHz respectively. The lower panel shows the evolution of the radio spectral index between 5 and 8.4 GHz (light red) and 8.4 and 22 GHz (dark red); open circles denote two epochs of near-simultaneous observations for the lower band index. While the X-rays have shown considerable variability during the years since the late 2017 CL event, the radio remained quiescent in all bands until exhibiting an exponential rise over a few months in early 2023. The radio evolution since has shown only mild variability at or slightly below the peak radio flux density reached in June 2023, with the exception of K-band.}
    \label{fig:lightcurve}
\end{figure*}

The resulting total intensity map of 1ES~1927+654 shows an unresolved point source of 53 mJy (reference frequency 5.5 GHz) with very low to zero polarization. The flux density is consistent with the VLBA flux density measurement of 49 mJy at 4.98 GHz on the same date given typical uncertainties on the absolute flux density scale of $\sim 5$--10\%. The linear polarization intensity map formed from the Stokes Q and U images is shown in Figure~\ref{fig:vla} and shows a point source with peak flux density 0.29 mJy, which is approximately 7~times the (polarized intensity) image RMS. However, similar low-level peaks which are clearly noise can be seen in several locations in the image, with peak levels up to $\sim$0.1 mJy. The peak of polarized intensity is also not perfectly aligned with the peak in the Stokes' I image. Thus some caution is warranted and we interpret this as an upper limit of 0.6\% on the linear polarization fraction at C-band. The Stokes~V image does not show any signs of an excess and based on the V-band RMS we report an upper limit of 0.2\% on the circular polarization.

\subsection{Arcminute Microkelvin Imager (AMI) observations at 15.5 GHz}

1ES 1927+654 was observed on many occasions between 2023 Jun~07 and 2024 April~02 with the Large Array of Arcminute MicroKelvin Imager (AMI; \citealt{2008MNRAS.391.1545Z, 2018MNRAS.475.5677H}). AMI consists of eight 12.8-m antennas sited at the Mullard Radio Astronomy Observatory near Cambridge, UK. The AMI receivers cover the band from 13 to 18 GHz, and are of a single linear polarisation, Stokes I+Q. We report here 58 observations taken over the above timeframe with a mean spacing of $\sim 5.3$ days. Analysis was done using custom software, \textsc{reduce\_dc} \citep{2013MNRAS.429.3330P}. Each observation consisted of multiple 10-min scans of 1ES 1927+654, interleaved with short ($\sim 2$-min) observations of a nearby compact source, which were used for phase calibration. The flux density scale was set using nearby observations of 3C286, which were usually made daily. The number of antennas available varied between observations, due to technical issues, and usually longer observations were made when there were fewer working antennas available. The day-to-day flux density uncertainty is estimated at $\sim 5$\%.

\subsection{e-MERLIN}

Regular observations from e-MERLIN cycle 16 program CY16025 were obtained as ``filler'' scans at C-band inserted into other accepted programs using that frequency. Partial results from this program, namely data from September to November 2023 so far have been made available as fully calibrated uvfits files. With only modest ``snapshot'' depth observations the UV coverage is not always sufficient for reliable imaging deconvolution, so we have opted for a non-imaging analysis method to obtain the source flux in the e-MERLIN observations. In particular, we use the CASA task \texttt{uvmodelfilt} to fit a simple point-source model to the visibilities, adjusting the starting values to ensure stability and convergence. The model column of the MS (measurement set) file is then populated according to the fit results and we apply standard self-calibration cycles with decreasing solution intervals to then improve the applied calibration, re-running the model fit at each step. We conducted several rounds of phase-only (non-cumulative) self-calibration followed by a single round of amplitude and phase calibration. In the middle panel of Figure~\ref{fig:lightcurve}, we show the preliminary results from this program, as 3~data points corresponding to roughly weekly averages.

\subsection{Submillimeter Array (SMA)}

The source 1ES~1927+654 was observed twice by the SMA in Hawaii, on 31 July 2023 and 11 June 2024 (hereafter epoch 1 and 2). In both observations there were six SMA antennas operating in similar compact configurations with baselines up to $\sim$80 m. In epoch 1 the target was observed in a single frequency band using two orthogonally polarized double sideband receiver sets, each processing 12 GHz of bandwidth per sideband (providing a total of 48 GHz of processed bandwidth) centered at a frequency of 225.5 GHz ($\lambda \sim$1.35 mm). In epoch 2 the observation included simultaneous dual band observations, each with a single polarization (providing 24 GHz of processed bandwidth each), one again centered at 225.5 GHz and the other at 347 GHz ($\lambda \sim$870 $\mu$m.  The weather was very good in both observations, with the water vapor column ranging between 1.5 and 2 mm pwv. Observations of 1ES~1927+654 were interleaved with observations of nearby gain calibration sources, J1806+698 and J1927+739, with the absolute flux scale calibrated primarily against the continuum of MWC349A, with added checking in epoch 1 using Ceres and Callisto. Standard reduction using the SMA MIR calibration suite was performed for both epoch 1 and 2.\footnote{https://lweb.cfa.harvard.edu/rtdc/SMAdata/process/mir/} The data for epoch 2 were also processed using the COMPASS pipeline reduction package which provides several improved and automated steps for data flagging and quality control (Keating, private communication). The total usable on-source integration time was 7.13 hours and 6.97 hours for epochs 1 and 2, respectively. The synthesized spatial resolution at 225.5 GHz was similar for both epochs, $\sim$3.''$\times$3.0'', while the higher frequency band in epoch 2 at 347 GHz achieved a resolution of  2.2''$\times$1.9''. For epoch 1 (225.5 GHz), the flux density was found to be 5.49$\pm$0.29 mJy, determined from a vector average of the calibrated visibility data, and confirmed using imaging using the AIPS task \texttt{IMAGR}. For epoch 2, the flux density at 225.5 GHz was measured to be 5.73$\pm$0.35 mJy (consistent with epoch 1), and at 347 GHz  the flux density was 4.57$\pm$1.27 mJy using the COMPASS pipeline reduction and imaging path.

\begin{figure*}[!t]
    \centering
    \includegraphics[width=0.95\textwidth]{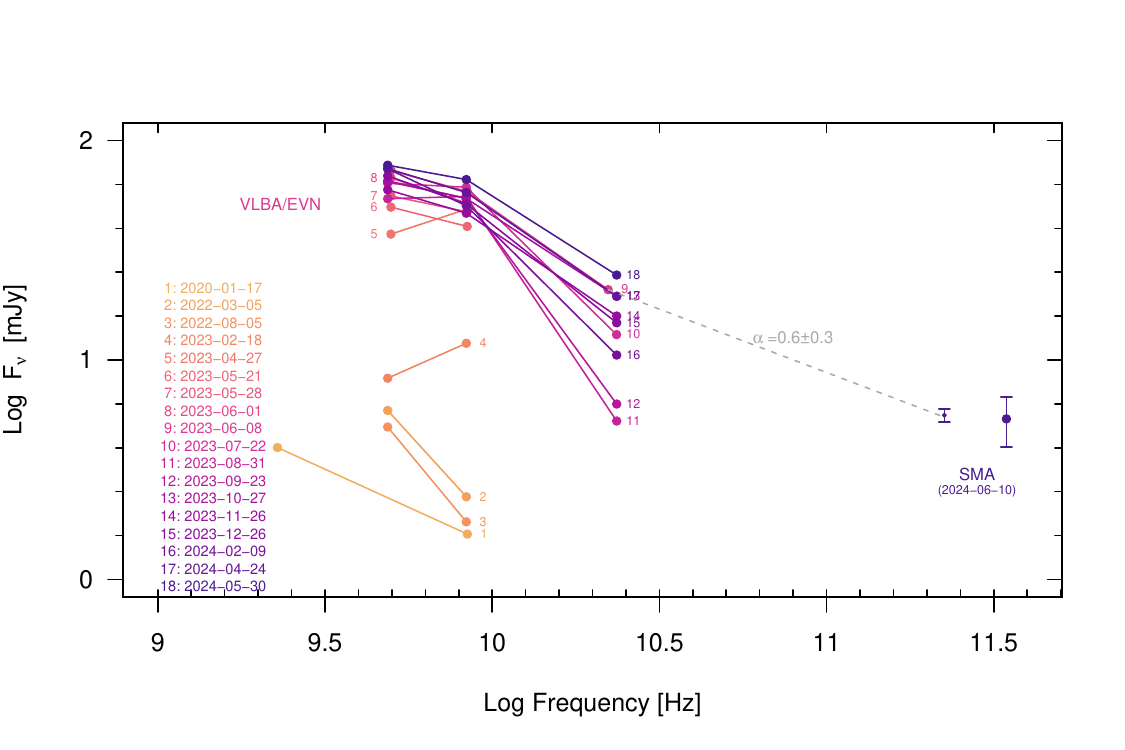}
    \caption{Plots of the radio spectral energy distribution (VLBI and SMA only), from 2020 January through 2024 April.  As shown, the source exhibited a flat/hard spectrum during the initial rise in early 2023, but since then shows a fairly consistent GHz-peaked spectrum. The peak of the spectrum appears to be at or just below 5 GHz. The SMA observations shown were taken on 10 June 2024 at 225 and 345 GHz and suggest a flat spectrum ($\alpha=0.1\pm0.6$) in the sub-mm band, possibly consistent with a coronal origin. The 225 GHz flux in June 2024 is consistent with the value obtained in July 2023 at the same frequency. }
    \label{fig:radio_SED}
\end{figure*}

\section{Results and Discussion} \label{sec:result}

\subsection{Radio Lightcurve and SED evolution} \label{sec:lightcurve}

In Figure~\ref{fig:lightcurve} we show the main result for 1ES~1927+654, which is a $\sim 40-$ and 60-fold increase at 5 and 8.4 GHz in the core radio flux density over a very short time period from January to June 2023. The upper panel shows the 0.3--2 keV soft X-ray flux as seen by \emph{Swift}/XRT for comparison (see \citet{laha2024} for further details). The central panel shows the total radio flux in the VLBA observations at 5.4, 8.4, and 22.2/23.5 GHz (C, X, and K-band) as well as fluxes from lower-resolution instruments e-MERLIN (5 GHz) and AMI (15.5 GHz). The 5--8.4 GHz and 8.4--23 GHz spectral indices are shown in the lower panel over the same timeframe. The peak radio flux was likely reached between the 8~June and 22~July epochs of VLBA monitoring, in agreement with the peak in the AMI lightcurve on 27~June. The 5.4--8.4 GHz spectral index became notably ``flat'' ($\alpha<0.5$ for $\nu^{-\alpha}$) during the fast rise in early 2023, and has remained so in nearly all epochs in the year following.  The radio loudness is now $R\approx100$, where $R$ the ratio of the optical (traditionally B-band) flux to that in the radio at 5 GHz. We use the recent consistent quiescent optical magnitude of m$_g$=17 as seen by the Zwicky Transient Facility \citep[ZTF][]{masci2019}, equivalent to a flux of $\sim$0.6 mJy for the optical flux density. The source can thus be argued to now be radio-loud, though the dividing line between radio-quiet, radio-intermediate, and radio-loud is debated \citep[e.g.][]{falcke1996}

\begin{figure*}
    \centering
    \includegraphics[width=\textwidth]{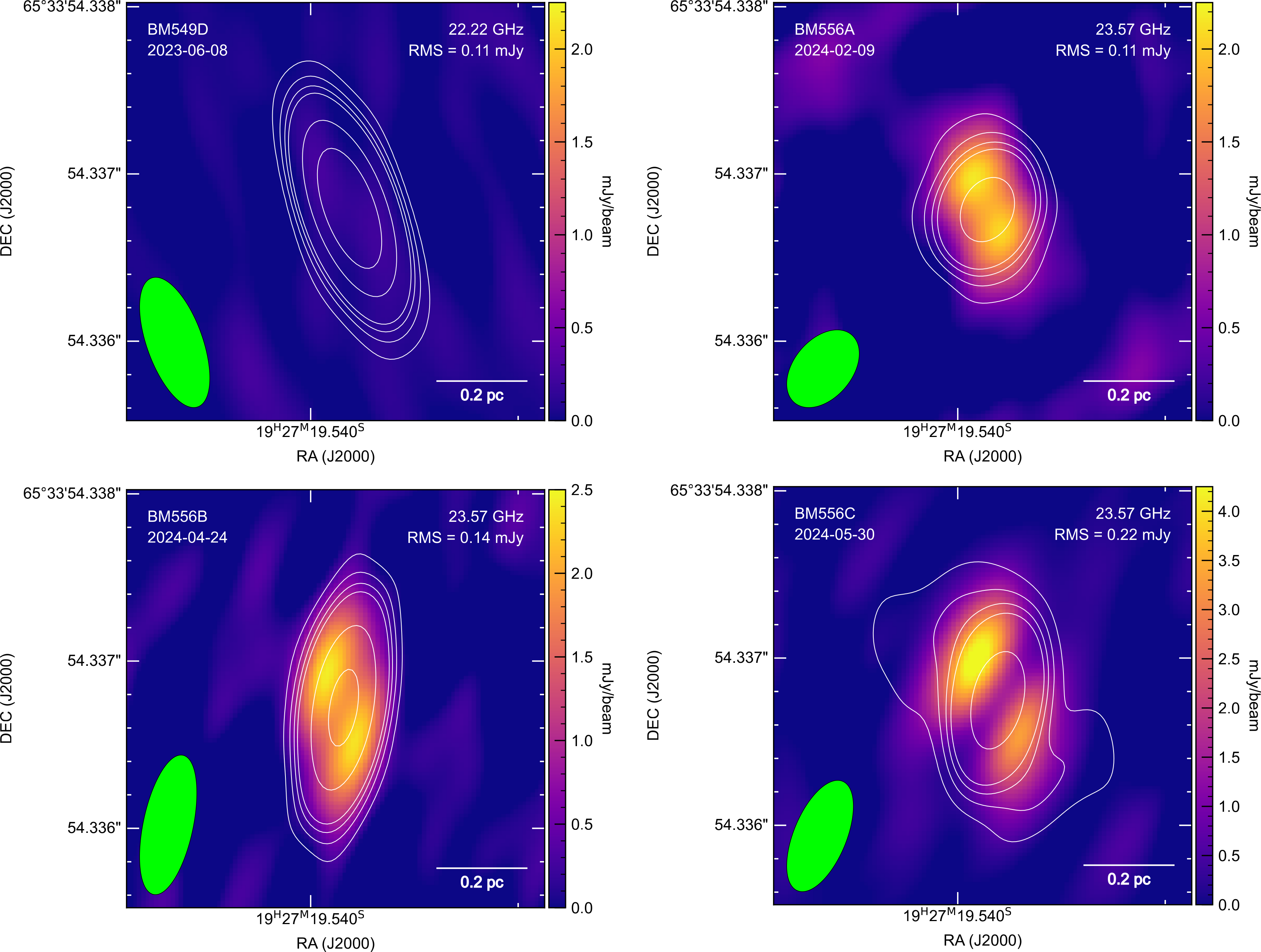}
    \caption{Core-subtracted K-band VLBA images of 1ES 1927+654 on 8 June 2023 (top left) 9 Feb 2024 (top right), 24 Apr 2024 (bottom left) and 30 May 2024 (bottom right). The contours correspond to the non-core-subtracted intensity and are drawn at 5, 10, 15, 20 and 50 times the non-core-subtracted image RMS (given at upper right in each panel).  The longer February 2024 observation resulted in better UV coverage and a less elongated synthesized beam (green oval at lower left in each panel) than the other observations. The extended structure apparent in early- to mid-2024 is entirely absent at the time of the initial radio outburst in June 2023.}
    \label{fig:ksub}
\end{figure*}
The near-weekly cadence of VLBA observations in May-June 2023 fortunately caught the source during the exponential rise phase. A fit to the six C-band observations from 18 February to 8 June 2023 with an exponential function gives a characteristic timescale of $\sim$ 44 days, with an equivalent light-crossing size limit of about 0.04 pc (assuming no significant beaming). This is just below the resolution of our K-band VLBA imaging which has a synthesized beam on the order of 0.4--0.8 mas or 0.15--0.30 pc.

In the nearly 1~year since the radio outburst began, the source has maintained near-peak radio luminosity with only low-level variability from 5--15.5 GHz.\footnote{ The 22.2-23.6 GHz K-band flux shows a somewhat higher degree of variability, but  we strongly suspect that some of the early K-band observations with quite low fluxes compared to the rest were affected by decoherence losses due to a less ideal phase calibrator (BM550A-D observations in particular).  In some of these epochs we also were not able to self-calibrated at K-band. We adopt a larger error bar of 30\% vs. a standard 10\% for the non-self-calibrated epochs in Figure~\ref{fig:lightcurve}).}The average 5.4 GHz flux density of 65 mJy since the peak in 2023 June corresponds to a radio power of $\nu L_{\nu,\mathrm{R}}=2.6\times10^{39}$ erg\,s$^{-1}$.

The VLBI-resolution radio spectral energy distribution from 2020 January to 2024 April (log\,$F_\nu$ vs log\,$\nu$) is shown in Figure~\ref{fig:radio_SED}, where we plot the total flux density. The spectrum is notably curved with a peak at or just below 5 GHz with an appearance similar to Gigahertz-peaked spectrum (GPS) AGN sources, as further discussed below. We also include in this figure SMA observations from June 2024, at 225 and 345 GHz (the 2023 July observation at 225 GHz only resulted in an almost identical flux value to that of 2024 and is not shown). Both the sub-mm band SMA 225-345 GHz spectral index ($\alpha=0.1\pm0.6$) and that between 23.6-225 GHz ($\alpha=0.58\pm0.28$) suggests spectral hardening compared to the much steeper spectrum at the same epoch measured between 8.4 and 23.6 GHz ($\alpha=1.05\pm0.26$). While the SMA observations are much lower resolution and could include a contribution from arcsecond-scale emission (i.e. cold dust), the nearly flat sub-mm spectrum is consistent with the observations and expectations of synchrotron emission from a compact corona \citep[e.g.]{raginski2016} rather than the Rayleigh-Jeans tail of a cold dust component. Additional higher-frequency observations should help clarify the degree of jet contribution to the SMA band and better constrain a possible turnover above 300 GHz.

\subsection{Extended Source Structure and Proper Motion}
\label{ext}
VLBA imaging of 1ES~1927+654 has previously shown a low-level resolved component on $\sim$few pc scales with a typical flux of no more than a few mJy. As previously presented in \cite{Laha_2022ApJ...931....5L}, the C-band (5.4 GHz) VLBA image of the source taken in 2021 March shows a central peak of approximately 2 mJy\,beam$^{-1}$ while the total flux is 5.5±0.5 mJy (indicating a resolved component). Fitting of the visibilities with a modified version of DIFMAP \citep{agniva_thesis,DIFMAP} showed that the extended flux was consistent with a disk-like component of uniform surface brightness (total flux $\sim 4$ mJy) and a size of $3.5 \times 4.5$ mas$^2$ ($1.3 \times 1.7$ pc$^2$). Similar results for the peak and extended component were obtained for the March and August 2022 observations,  with some hints of a reduction in the disk size (by $\sim$20\%) in the latter epoch \citep{Ghosh_2023ApJ...955....3G}. Surprisingly, the extended component was not detected at all (size $<0.1$ pc, flux $<0.1$ mJy) at C-band by 27 April 2023, when the unresolved core had begun to rise. It is possible that this emission arose from the slowly-expanding result of a previous outburst which became optically thin and went below the level of detection by early 2023; unfortunately we lack multi-band observations from that time for a more clear spectral analysis.

While a detailed investigation of the source structure through visibility-fitting of all VLBA epochs (as well as simulations to better estimate the dominant systematic contribution to the errors) is deferred to a future publication, we show in Figure~\ref{fig:ksub} core-subtracted K-band VLBA images from 8~June 2023 at 22.2 GHz (epoch BM549D) and 9~February, 24~April 2024, and 30 May at 23.6 GHz.  The February 2024 observation is deeper due to combining the alotted time of two observations from a nominally monthly monitoring program (7~hr vs 3.5~hr scheduling block) while the May observation was augmented to make a 6.2 hr scheduling block; the UV coverage in February and May is therefore better than in April, and this is likely the reason for the larger errors on component fits for the April epoch.

\begin{deluxetable*}{cccccccc}
\tabletypesize{\small}
%\tablenum{1}
\tablecaption{Results of DIFMAP fitting to resolved components in K-band imaging \label{tab:difmap}}
\tablehead{
\colhead{Observation} & \colhead{Decimal Date}& \colhead{Component} & \colhead{Flux$^\dagger$} & \colhead{$\delta$ East}  & \colhead{$\delta$ North} & \colhead{Separation} & \colhead{P.A.}\\
\colhead{} & \colhead{} &\colhead{} &  \colhead{(mJy)}& \colhead{($\mu$as)}& \colhead{($\mu$as)}& \colhead{(mas)}& \colhead{($\degree$)}\\
}
\colnumbers
\startdata
BM556A & 2024.107 & core & 7.66&\nodata & \nodata& \nodata & \nodata\\
       &          & NE & 1.88$\pm$0.03  & 88$\pm$4&  218$\pm$5 & 0.40$\pm$0.01 & 23$\pm$1\\
       &          & SW & 1.87$\pm$0.03 &  $-$70$\pm$4 &  $-$146$\pm$5 & & \\[1ex]
BM556B & 2024.311 & core & 15.1 & \nodata & \nodata & \nodata & \nodata\\
       &          & NE & 2.26$\pm$0.05  & 135$\pm$7& 173$\pm$32&  0.45$\pm$0.05& 24$\pm$4\\
       &          & SW & 2.23$\pm$0.05  &  $-$47$\pm$7 & $-$242$\pm$32 & \\[1ex]
BM556C & 2024.41 & core & 14.0 & \nodata & \nodata& \nodata & \nodata\\
       &         &  NE & 4.70$\pm$0.05  & 164$\pm$2& 284$\pm$3& 0.50$\pm$0.01& 31$\pm$1\\
       &         & SW & 3.51$\pm$0.03  & $-$93$\pm$2 & $-$142$\pm$3 & \\
\enddata
\tablecomments{$\dagger$ The flux listed for the core in each epoch is the value used in the core subtraction. }
\tablecomments{The quoted errors are the formal 1$\sigma$ values from the fitting and do not account for systematic effects.}
\end{deluxetable*}
After core subtraction, two peaks on opposite sides of the core are apparent in the residual images. A robust kinematic analysis is not possible at this time due to the limited number of epochs available, but we do report the results from directly fitting the visibilities for the Feb, Apr, and May 2024 K-band images in Table~2 with the caveat that the errors are underestimated due to unaccounted systematic effects \citep[see e.g.][]{pushkarev2012}.\footnote{Estimates of the systematic contribution to the error requires either extensive simulations of the data or time to accumulate many additional epochs of observation -- both will be presented in a future paper. The EVN K-band observations, though reaching comparable theoretical sensitivity to the VLBA due to long total observing times, lack the short-timescale signal/noise required for self-calibration and have large gaps in UV coverage on the scale of the extended components and therefore do not detect them. However \es is approved for continued VLBA monitoring with deeper K-band observations through 2025.} The components were fit as point sources using the DIFMAP package, and the resulting flux density and position relative to the core are given in columns 4-6. In column 7 we give the total distance from the core position and the associated position angle from the core.

While the fits appear to suggest both increasing flux and separation in the resolved components (equivalent to an apparent separation speed of 0.36$c$, or half that for a one-sided jet), we again urge caution due to the under-estimated errors, until more epochs of observation become available. Taking the mean separation of the components (0.45 mas) and assuming the outflow began with the onset of the radio flare in February 2023 (14 month time baseline), also yields a similar separation speed estimate, of 0.38$c$. Additional epochs of observation with good UV coverage are needed to confirm these initial estimates. Interestingly, the above speed estimates match the outflow speed (of 0.2$c$) inferred from the 1 keV emission line seen during the super-Eddington state of 2020-2021 -- a time when radio observations were unfortunately lacking \citep{masterson2022}.  If the extended resolved features represent a jet-driven outflow, continued observations of the source should more clearly resolve these components with time.

The orientation of \es\, has not been clearly determined, starting from its early designation as a `naked' or true type II AGN \citep{boller2003}, where the lack of obscuration seemed to point to an unusually under-luminous broad line region rather than greater obscuration consistent with standard AGN unification. Arguments have been made for both a large \cite[e.g. $\sim$85$\degree$ based on Compton reflection model fits as in][]{gallo2013} and small orientation angles \citep[e.g.][]{Trakhtenbrot_2019ApJ...883...94T}. The flux and speed of the bipolar outflow components detected in VLBI imaging can in principle constrain the orientation angle of the AGN, assuming the jet/outflow is perpendicular to the disk and intrinsically symmetric in properties \citep[see discussion and equations in e.g.][]{meyer2018}. While additional epochs are needed to provide a robust orientation estimate, the flux ratio of the jet/counterjet of $\sim$1 in combination with the low but still mildly relativistic apparent advance speed can at least rule out an orientation angle $\lesssim$ 20$\degree$.

\subsection{Source of the radio emission}

Radio emission in radio-quiet AGN can arise from a number of sources including star formation and shocks from AGN-driven winds, small-scale jets, and/or the same compact corona that gives rise to X-ray emission \citep[see e.g.][for a recent review]{panessa2019}. In \es, the first two origins can be ruled out easily based on both variability and physical scale. Indeed, comparison of low-resolution VLA and AMI observations taken at the same time as our VLBA observations shows no significant difference in total flux as might be expected if radio emission on larger scales (which would be resolved out by the VLBA) were significant. A strictly coronal origin for the GHz band compact emission is also precluded based on the requirements inferred from the X-ray observations \citep{laha2024}, i.e. the very small size ($\sim$ 10 $r_g$) and high magnetic field strength (on the order of $10^4-10^5$ G). Besides very strong synchrotron self-absorption (SSA) which would be expected in such a source, there will be no significant synchrotron emission below the cyclotron frequency of $\sim$(30 GHz)(B/1e4 G). 

A small-scale synchrotron jet/outflow\footnote{In this letter we use the terms jet and outflow interchangeably with regard to the resolved emission without intending them to represent different scenarios. } thus appears most consistent with the radio emission, even before the recent radio outburst.  In this case the unresolved radio emission we observe arises on much larger scales than the X-ray emitting corona. Taking a flux of 45 mJy at 15 GHz and radio spectral index $\alpha=1$ as typical values since the flare began, and assuming a source size of 0.05 pc, a minimum Lorentz factor $\gamma=1$ and no beaming, we use the minimum-energy condition (equipartition) to derive a magnetic field for emitting region of $B$=0.3~G \cite[standard formulae can be found in e.g.][]{Longair_book}.

For a synchrotron origin, the lack of polarization for \es in the 5 GHz VLA observation does not necessarily cause a problem, since internal Faraday depolarization is expected at the self-absorption frequency unless the plasma is dominated by highly relativistic electrons, with a lower-bound cutoff of $\gamma_\mathrm{min} \sim 10^2$ \citep{jones1977a}, as is likely the case in more powerful jets (i.e.\ blazars). Indeed, high rotation measures and low observed radio polarization are  typical of GPS sources \citep{odea1998}. 

While a full exploration of the X-ray and other multiwavelength observations of this source are given in the companion publication by \cite{laha2024}, we can make some brief comments here. While a significant rise in the (0.3-2 keV) soft X-ray band is apparent in Fig.~\ref{fig:lightcurve}, the hard X-rays have risen only slightly by a factor of perhaps 2-3, and the hardness ratio is now less than half what it was before the soft X-ray rise. Similarly, the X-ray spectral index went from typical values of 2.6-2.8 prior to the radio flare to 3.2-3.3 after. The soft X-ray emission can be fit with a thermal component with consistently low temperature (regardless of normalization) of $kT =0.13-0.16$ keV.

Interpreting these observations, we first note that the X-ray to radio luminosity ($10^{4}$) is far too high for both to arise from nonthermal processes in a very compact jet (i.e. inverse Compton emission) as is likely occuring in Compact Symmetric Objects \citep[CSO,][]{malgosia2023}, which this source otherwise resembles. Using the equipartition value of the magnetic field and assuming a `light' jet (electron-positron), and sub-relativistic speed of 0.3c, we estimate a minimum total kinetic power for the jet of $L_\mathrm{kin,min}$=$10^{43}$erg\,$s^{-1}$, which is comparable to the observed power radiated in soft X-rays as of mid-2024. (Here we use $L_\mathrm{kin,min}$=\,$\pi\,R^2\,(\Gamma^2 \beta)\,c\,U_B$ with $U_B$=0.0045~erg~cm$^{-3}$, $\Gamma^2 \beta \approx1$) The onset of the soft X-ray rise just before the radio, as well as the spectral shape in the soft X-rays, could be consistent with shocked gas being driven by the newly launched radio jets, as an alternative to a coronal origin. In either case, assuming the X-ray and radio emission are related, the delay in the radio emission could be explained by the presence of screening material  causing free-free absorption; alternatively, the jet/outflow may be either too compact or simply inefficient at particle acceleration in the initial phase. If there is an external screen of hot gas, it must be on a larger scale than the jet-heated gas, on the order of the size of the broad line region (30 light-days or 0.25 pc), to account for the $\sim$200 day delay \citep[see][]{laha2024}. 

\subsection{Comparison with young/short-lived jet classes}
As noted, the current spectrum of 1ES~1927+654 bears a resemblance to GPS sources, which are young radio jets less than 1000 years old and usually  $<$1 kpc in extent \citep[e.g.][]{odea2021}. The compact radio emission is due to synchrotron with the characteristic GHz-frequency peak due to SSA or free-free absorption. The advance speeds of these jets are typically low, on the order of 0.1--0.2$c$. A related class of sources are known as compact symmetric objects or CSO, which are double-lobed radio sources $<$1 kpc in extent \citep{readhead1994,readhead1996}. Unlike GPS sources, the CSO class is primarily a morphological one and requires high-resolution radio imaging for identification. They are often similarly identified as young radio sources $\lesssim 1000$ yr in age and have relatively slow rates of growth in size, of $\sim$0.3$c$ on average \citep{taylor_kinematic_2000,an2012}.  In terms of their radio spectra, most CSO sources (e.g. 82\% in the study of \citealt{tremblay_compact_2016}) exhibit a GPS-like spectrum, and can be core or lobe-dominated (type 1 or 2, respectively). Very recent work on the kinematics of a carefully selected sample of mostly CSO-2 suggests that they should be considered a distinct class of `short-lived' jets rather than simply young versions of more classical powerful radio-loud AGN \citep{kiehlmann_compact_2023}. This fits well with the idea that CSO-type jet activity is powered by a ``single fueling event'', e.g. the disruption of a single star by the central black hole \citep{readhead1994,an2012,kiehlmann_compact_2024}.% -- an idea we will return to in the next section focusing on comparisons to TDE.

Although direct evidence of newly formed radio jets in AGN are rare, it is not unprecedented. A recent study comparing the VLA All-Sky Survey (VLASS) epoch~1 observations (2017--2019) to the earlier FIRST (Faint Images of the Radio Sky at Twenty cm; 1993–2011) survey discovered 14 new radio-loud sources which have turned on sometime in the last $\sim$20 yr, and these also have peaked radio spectra resembling GPS sources \citep{nyland2020}. The radio emission was roughly constant on few-month timescales on VLA follow-up, and the typically tens-of-GHz frequencies of the radio spectral peaks appear consistent with the expected size$-$peak frequency relation of young jets \citep[][]{odea1998,jeyakumar2016}.  The same is not true of 1ES~1927+654, where we clearly see the spectral peak is below 5 GHz, contrary to expectations for a very young jet. On the other hand its low luminosity does match its small size ($<1$ pc) according to the CSO `HR diagram' of \cite{readhead2024}, where only one other CSO is within an order of magnitude of the very low ($\sim10^{22}$ W/Hz) radio luminosity of \es.

While there is clear similarity between the case of \es\, and known young/short-lived jets, the radio powers of classical CSO and GPS sources are generally much higher than 1ES~1927+654, by a factor of $>$100 \citep[][]
{odea1998, odea2021}. Even for the recently discovered VLASS sources, which extend down to lower radio power ($\log L_\mathrm{3\,GHz}=40.5$) than classical GPS samples, the lowest is still an order of magnitude more powerful in the radio than 1ES~1927+654. However this may simply reflect the ``down-sized'' nature of 1ES~1927+654 which likely hosts a central black hole of only $\sim10^6\,M_\odot$ \citep{li2022}.  While there are some indications that GPS sources have somewhat smaller black hole masses on average than their classical radio-loud QSO cousins \citep[e.g.][]{gu2009,wu2009}, typical values are on the order of $M_\mathrm{BH}\sim 10^8\,M_\odot$, or 100x that of \es. It is plausible that the lack of lower-power CSO sources in our current samples may simply be a selection effect. Indeed, although focusing on the far more prevalent non-CSO jet classes, the LOFAR Two-Metre Sky Survey found that the morphological class of edge-brightened (FRII) type jets, for decades thought to occur only at relatively high luminosities, actually extend down to luminosities 3 orders of magnitude below the traditional Fanaroff \& Riley division \citep{mingo2019}. As CSO sources appear to be relatively rare (6-8\%) in radio-selected AGN samples, and only recently have moderately sized complete samples been compiled \citep[e.g. 79 sources in][]{kiehlmann_compact_2024}, deeper surveys may well uncover more objects like \es.

\subsection{Comparison to late-time radio flares in TDEs}

One observation that bears further exploration is the initial exponential rise in radio flux for \es, which is in contrast with the prediction of a much slower $P\propto t^{2/5}$ behavior predicted for the turn-on of GPS sources \citep{an2012}.  Fast-rising radio brightening has been seen in TDEs, however \citep[e.g.][]{cendes_mildly_2022}, and the possibility that the 2017-2018 flare event in \es\, was a TDE is still open \citep{ricci2021}.TDE in existing AGN are theoretically expected \citep[e.g.][]{chan_tidal_2019} but are challenging to convincingly detect due to competition from normal AGN flares \citep[e.g.][]{auchettl2018}. Interestingly, some recent theoretical work has suggested that TDE in existing AGN may precipitate changing-look events \citep{wang2024}.

TDEs are seen to produce prompt radio emission (within $\sim$100 days) in only about 20-30\% of cases. The radio is typically attributed to synchrotron emission from the interaction of outflows with circum-nuclear material, as well as shocks formed in debris-stream collisions or, more rarely, relativistic jets \citep[see][for a recent review]{alexander2020}. However, very recent work shows that late-time radio activity actually occurs in $>$40\% of TDE with initial non- or low-level detections (lower limit being due to limited follow-up), with onset times ranging from a few hundred to $>$1000 days \citep[][and references therein]{cendes_ubiquitous_2023}. These can exhibit very fast rise timescales, e.g. faster than $F_\nu \propto t^5$ in AT2018hyz \citep{cendes_mildly_2022}. The latter case may be consistent with an off-axis relativistic jet, where the radio emission is initially `beamed away' before the jet decelerates enough to become visible \citep[][but see also \citealt{cendes_mildly_2022}]{sfaradi_off-axis_2023}. However misaligned jets cannot explain the majority of cases, most directly because of the high rate of on-axis jetted TDE implied, which is not observed, and also because a decelerating jet scenario typically predicts a radio peak on the order of 100 days after onset, far less than the observed values of $\gtrsim$ 700 days \citep{matsumoto_generalized_2023,cendes_ubiquitous_2023,beniamini2023}.  The origin of the probable outflows responsible for late-time radio emission in TDE is still under discussion, however  \cite{matsumoto_late-time_2024} are able to explain delays up to those observed ($\sim10^3$ days) as a natural consequence of non-relativistic to very mildly relativistic outflows ($\beta<0.15$) and the effects of a flattening density profile of the circum-nuclear material.

The case of \es\, does show some similarities to TDE with late-time radio emission. The timescale is only slightly longer than the most delayed onset in TDE seen so far (at $\sim 1800$ days), though one of the key conclusions of the recent discovery of late-time emission is that further long-term follow-up, possibly on decade timescales, is needed to fully characterize the phenomenon \citep{cendes_ubiquitous_2023}.
\cite{alexander2020} also note that a significant outflow may only be launched after (and if) the accretion onto the BH reaches a super-Eddington phase. For \es\, the time delay of the radio onset from its superluminal period in 2020 is $\sim$1000 days. 

If our initial estimate of an outflow of speed $\sim$0.3$c$ stands, then the outflow in \es\, is in an interesting middle regime between the sub-relativistic outflows inferred in the sample of late-time TDE radio sources and the relativistic jets seen in small fractions of both AGN and TDE. The radio luminosity of \es\, is comparable to or slightly above the upper end of what is seen for non-relativistic TDE outflows (typically $<10^{39}$ erg/s) but well below that of most radio-loud AGN or e.g., the famous jetted TDE Swift J1644+57 \citep{berger2012}.  
The BH mass of \es\, may explain some of the similarity in energy scale, as the BH masses for TDE tend to be substantially smaller than in AGN on average \citep[e.g., $5\times 10^5 - 10^7\,M_\odot$ in the sample of][]{ryu2020} reflecting the population of quiescent BH and the fact that disruption events around BH larger than $\sim10^8\,M_\odot$ do not produce observable radiation. However, a down-sized black hole does not mean that highly relativistic jets cannot form, as clearly demonstrated by the recent discoveries of relativistic, gamma-ray emitting jets in narrow-line Seyfert galaxies \citep[e.g.][]{foschini2015} as well as low-power but highly collimated/fast FR II jets, mentioned previously \citep{mingo2019}. It seems likely that the difference comes down to the nature of jet-launching itself, and further study of sources like \es\, will be key to advancing our understanding.

\section{Conclusions}

A little over 5~years after it became one of extremely few AGN with a directly-observed changing-look event \citep{Trakhtenbrot_2019ApJ...883...94T}, \es\, has recently exhibited a significant radio brightening consistent with a newly-launched radio-emitting outflow, approximately 1800 days after the initial CL event.  A fortuitously timed weekly VLBA monitoring program in May to June 2023 caught the exponential rise and peak of the radio onset, which reached a level of 40 and 60 times previous radio core luminosities at 5 and 8.4 GHz, respectively. The source has maintained a relatively steady radio emission with a spectrum reminiscent of gigahertz-peaked sources, for over 1 year without obvious signs of fading or further increase.  The soft X-ray emission, which began rising a few months prior to the flare, may arise from shock-heated gas impacted by the jet; in this scenario the radio jet may have been first screened by larger-scale and pre-existing hot gas before breaking through, consistent with the delayed but rapid radio brightening. Our most recent high-resolution VLBA imaging at 23.6 GHz shows bipolar radio extensions of similar brightness separated by approximately 0.45 mas or 0.15 pc, with a tentative expansion of separation speed of $\sim$0.3$c$. The resolved structures resemble a  very low-luminosity compact symmetric object (CSO), which have recently been suggested to be powered by TDE events \citep{readhead2024}. The radio outflow characteristics also bear some resemblance to those TDE with late-time radio emission, though as noted in previous works, the X-ray properties of this source are considerbly different from a TDE, likely due to the presence of a pre-existing accretion disk.  Continued follow-up with high-resolution and multi-frequency radio observations will allow us to further constrain the kinematics and energetics of the outflow in this ever-changing and unique AGN.

\begin{acknowledgments}

The National Radio Astronomy Observatory is a facility of the National Science Foundation operated under cooperative agreement by Associated Universities, Inc.  This work made use of the Swinburne University of Technology software correlator, developed as part of the Australian Major National Research Facilities Programme and operated under licence.

e-MERLIN is a National Facility operated by the University of Manchester at Jodrell Bank Observatory on behalf of STFC, part of UK Research and Innovation.

We thank the Mullard Radio Astronomy Observatory staff for the AMI observations. 

We acknowledge Phil Cigan and the MultiColorFits package which was used to create the image figures \citep{cigan2019}.

MN is supported by the European Research Council (ERC) under the European Union’s Horizon 2020 research and innovation programme (grant agreement No.~948381) and by UK Space Agency Grant No.~ST/Y000692/1.

The Submillimeter Array is a joint project between the Smithsonian Astrophysical Observatory and the Academia Sinica Institute of Astronomy and Astrophysics and is funded by the Smithsonian Institution and the Academia Sinica. Maunakea,
the location of the SMA, is a culturally important site
for the indigenous Hawaiian people; we are privileged
to study the cosmos from near its summit.

FP acknowledges
financial support from the Bando Ricerca Fondamentale INAF 2023. 

Based on observations obtained with the Samuel Oschin Telescope 48-inch and the 60-inch Telescope at the Palomar
Observatory as part of the Zwicky Transient Facility project. ZTF is supported by the National Science Foundation under Grant
No. AST-2034437 and a collaboration including Caltech, IPAC, the Weizmann Institute for Science, the Oskar Klein Center at
Stockholm University, the University of Maryland, Deutsches Elektronen-Synchrotron and Humboldt University, the TANGO
Consortium of Taiwan, the University of Wisconsin at Milwaukee, Trinity College Dublin, Lawrence Livermore National
Laboratories, and IN2P3, France. Operations are conducted by COO, IPAC, and UW.

\end{acknowledgments}

\vspace{5mm}
\facilities{Very Large Array (VLA); Very Long Baseline Array (VLBA); European VLBI Network (EVN); Arcminute Microkelvin Imager (AMI); enhanced Multi Element Remotely Linked Interferometer Network (e-MERLIN); Submillimeter Array (SMA); Swift X-ray Observatory}

\bibliography{1ES1927_Radio_Outburst}{}

\begin{thebibliography}{}
\expandafter\ifx\csname natexlab\endcsname\relax\def\natexlab#1{#1}\fi
\providecommand{\url}[1]{\href{#1}{#1}}
\providecommand{\dodoi}[1]{doi:~\href{http://doi.org/#1}{\nolinkurl{#1}}}
\providecommand{\doeprint}[1]{\href{http://ascl.net/#1}{\nolinkurl{http://ascl.net/#1}}}
\providecommand{\doarXiv}[1]{\href{https://arxiv.org/abs/#1}{\nolinkurl{https://arxiv.org/abs/#1}}}

\bibitem[{{Alexander} {et~al.}(2020){Alexander}, {van Velzen}, {Horesh}, \& {Zauderer}}]{alexander2020}
{Alexander}, K.~D., {van Velzen}, S., {Horesh}, A., \& {Zauderer}, B.~A. 2020, \ssr, 216, 81, \dodoi{10.1007/s11214-020-00702-w}

\bibitem[{{An} \& {Baan}(2012)}]{an2012}
{An}, T., \& {Baan}, W.~A. 2012, \apj, 760, 77, \dodoi{10.1088/0004-637X/760/1/77}

\bibitem[{{Antonucci}(1993)}]{antonucci1993}
{Antonucci}, R. 1993, \araa, 31, 473, \dodoi{10.1146/annurev.aa.31.090193.002353}

\bibitem[{{Auchettl} {et~al.}(2018){Auchettl}, {Ramirez-Ruiz}, \& {Guillochon}}]{auchettl2018}
{Auchettl}, K., {Ramirez-Ruiz}, E., \& {Guillochon}, J. 2018, \apj, 852, 37, \dodoi{10.3847/1538-4357/aa9b7c}

\bibitem[{{Beniamini} {et~al.}(2023){Beniamini}, {Piran}, \& {Matsumoto}}]{beniamini2023}
{Beniamini}, P., {Piran}, T., \& {Matsumoto}, T. 2023, \mnras, 524, 1386, \dodoi{10.1093/mnras/stad1950}

\bibitem[{{Berger} {et~al.}(2012){Berger}, {Zauderer}, {Pooley}, {Soderberg}, {Sari}, {Brunthaler}, \& {Bietenholz}}]{berger2012}
{Berger}, E., {Zauderer}, A., {Pooley}, G.~G., {et~al.} 2012, \apj, 748, 36, \dodoi{10.1088/0004-637X/748/1/36}

\bibitem[{{Bianchi} {et~al.}(2012){Bianchi}, {Maiolino}, \& {Risaliti}}]{bianchi2012}
{Bianchi}, S., {Maiolino}, R., \& {Risaliti}, G. 2012, Advances in Astronomy, 2012, 782030, \dodoi{10.1155/2012/782030}

\bibitem[{{Boller} {et~al.}(2003){Boller}, {Voges}, {Dennefeld}, {Lehmann}, {Predehl}, {Burwitz}, {Perlman}, {Gallo}, {Papadakis}, \& {Anderson}}]{boller2003}
{Boller}, T., {Voges}, W., {Dennefeld}, M., {et~al.} 2003, \aap, 397, 557, \dodoi{10.1051/0004-6361:20021520}

\bibitem[{{CASA Team} {et~al.}(2022){CASA Team}, {Bean}, {Bhatnagar}, {Castro}, {Donovan Meyer}, {Emonts}, {Garcia}, {Garwood}, {Golap}, {Gonzalez Villalba}, {Harris}, {Hayashi}, {Hoskins}, {Hsieh}, {Jagannathan}, {Kawasaki}, {Keimpema}, {Kettenis}, {Lopez}, {Marvil}, {Masters}, {McNichols}, {Mehringer}, {Miel}, {Moellenbrock}, {Montesino}, {Nakazato}, {Ott}, {Petry}, {Pokorny}, {Raba}, {Rau}, {Schiebel}, {Schweighart}, {Sekhar}, {Shimada}, {Small}, {Steeb}, {Sugimoto}, {Suoranta}, {Tsutsumi}, {van Bemmel}, {Verkouter}, {Wells}, {Xiong}, {Szomoru}, {Griffith}, {Glendenning}, \& {Kern}}]{Casa_2022PASP..134k4501C}
{CASA Team}, {Bean}, B., {Bhatnagar}, S., {et~al.} 2022, \pasp, 134, 114501, \dodoi{10.1088/1538-3873/ac9642}

\bibitem[{Cendes {et~al.}(2022)Cendes, Berger, Alexander, Gomez, Hajela, Chornock, Laskar, Margutti, Metzger, Bietenholz, Brethauer, \& Wieringa}]{cendes_mildly_2022}
Cendes, Y., Berger, E., Alexander, K.~D., {et~al.} 2022, The Astrophysical Journal, 938, 28, \dodoi{10.3847/1538-4357/ac88d0}

\bibitem[{Cendes {et~al.}(2023)Cendes, Berger, Alexander, Chornock, Margutti, Metzger, Wieringa, Bietenholz, Hajela, Laskar, Stroh, \& Terreran}]{cendes_ubiquitous_2023}
---. 2023, Ubiquitous {Late} {Radio} {Emission} from {Tidal} {Disruption} {Events},  arXiv.
\newblock \url{http://arxiv.org/abs/2308.13595}

\bibitem[{Chan {et~al.}(2019)Chan, Piran, Krolik, \& Saban}]{chan_tidal_2019}
Chan, C.-H., Piran, T., Krolik, J.~H., \& Saban, D. 2019, The Astrophysical Journal, 881, 113, \dodoi{10.3847/1538-4357/ab2b40}

\bibitem[{{Cigan}(2019)}]{cigan2019}
{Cigan}, P. 2019, {MultiColorFits: Colorize and combine multiple fits images for visually aesthetic scientific plots}, Astrophysics Source Code Library, record ascl:1909.002

\bibitem[{{Falcke} {et~al.}(1996){Falcke}, {Sherwood}, \& {Patnaik}}]{falcke1996}
{Falcke}, H., {Sherwood}, W., \& {Patnaik}, A.~R. 1996, \apj, 471, 106, \dodoi{10.1086/177956}

\bibitem[{{Foschini} {et~al.}(2015){Foschini}, {Berton}, {Caccianiga}, {Ciroi}, {Cracco}, {Peterson}, {Angelakis}, {Braito}, {Fuhrmann}, {Gallo}, {Grupe}, {J{\"a}rvel{\"a}}, {Kaufmann}, {Komossa}, {Kovalev}, {L{\"a}hteenm{\"a}ki}, {Lisakov}, {Lister}, {Mathur}, {Richards}, {Romano}, {Sievers}, {Tagliaferri}, {Tammi}, {Tibolla}, {Tornikoski}, {Vercellone}, {La Mura}, {Maraschi}, \& {Rafanelli}}]{foschini2015}
{Foschini}, L., {Berton}, M., {Caccianiga}, A., {et~al.} 2015, \aap, 575, A13, \dodoi{10.1051/0004-6361/201424972}

\bibitem[{{Gallo} {et~al.}(2013){Gallo}, {MacMackin}, {Vasudevan}, {Cackett}, {Fabian}, \& {Panessa}}]{gallo2013}
{Gallo}, L.~C., {MacMackin}, C., {Vasudevan}, R., {et~al.} 2013, \mnras, 433, 421, \dodoi{10.1093/mnras/stt735}

\bibitem[{{Ghosh} {et~al.}(2023){Ghosh}, {Laha}, {Meyer}, {Roychowdhury}, {Yang}, {Acosta-Pulido}, {Rakshit}, {Pandey}, {Gonz{\'a}lez}, {Behar}, {Gallo}, {Panessa}, {Bianchi}, {La Franca}, {Scepi}, {Begelman}, {Longinotti}, {Lusso}, {Oates}, {Nicholl}, {Cenko}, {O'Connor}, {Hammerstein}, {Jose}, {Gab{\'a}nyi}, {Ricci}, \& {Chattopadhyay}}]{Ghosh_2023ApJ...955....3G}
{Ghosh}, R., {Laha}, S., {Meyer}, E., {et~al.} 2023, \apj, 955, 3, \dodoi{10.3847/1538-4357/aced92}

\bibitem[{{Gu} {et~al.}(2009){Gu}, {Pak}, \& {Ho}}]{gu2009}
{Gu}, M.~F., {Pak}, S., \& {Ho}, L.~C. 2009, Astronomische Nachrichten, 330, 253, \dodoi{10.1002/asna.200811169}

\bibitem[{{Hickish} {et~al.}(2018){Hickish}, {Razavi-Ghods}, {Perrott}, {Titterington}, {Carey}, {Scott}, {Grainge}, {Scaife}, {Alexander}, {Saunders}, {Crofts}, {Javid}, {Rumsey}, {Jin}, {Ely}, {Shaw}, {Northrop}, {Pooley}, {D'Alessandro}, {Doherty}, \& {Willatt}}]{2018MNRAS.475.5677H}
{Hickish}, J., {Razavi-Ghods}, N., {Perrott}, Y.~C., {et~al.} 2018, \mnras, 475, 5677, \dodoi{10.1093/mnras/sty074}

\bibitem[{{Janssen, M.} {et~al.}(2019){Janssen, M.}, {Goddi, C.}, {van Bemmel, I. M.}, {Kettenis, M.}, {Small, D.}, {Liuzzo, E.}, {Rygl, K.}, {Martí-Vidal, I.}, {Blackburn, L.}, {Wielgus, M.}, \& {Falcke, H.}}]{picard}
{Janssen, M.}, {Goddi, C.}, {van Bemmel, I. M.}, {et~al.} 2019, Astronomy \& Astrophysics, 626, A75, \dodoi{10.1051/0004-6361/201935181}

\bibitem[{{Jeyakumar}(2016)}]{jeyakumar2016}
{Jeyakumar}, S. 2016, \mnras, 458, 3786, \dodoi{10.1093/mnras/stw181}

\bibitem[{{Jones} \& {O'Dell}(1977)}]{jones1977a}
{Jones}, T.~W., \& {O'Dell}, S.~L. 1977, \apj, 214, 522, \dodoi{10.1086/155278}

\bibitem[{Kiehlmann {et~al.}(2023)Kiehlmann, Readhead, O'Neill, Wilkinson, Lister, Liodakis, Bruzewski, Pavlidou, Pearson, Sheldahl, Siemiginowska, Tassis, \& Taylor}]{kiehlmann_compact_2023}
Kiehlmann, S., Readhead, A. C.~S., O'Neill, S., {et~al.} 2023, Compact {Symmetric} {Objects} -- {II} {Confirmation} of a {Distinct} {Population} of {High}-{Luminosity} {Jetted} {Active} {Galaxies},  arXiv.
\newblock \url{http://arxiv.org/abs/2303.11359}

\bibitem[{Kiehlmann {et~al.}(2024)Kiehlmann, Lister, Readhead, Liodakis, O’Neill, Pearson, Sheldahl, Siemiginowska, Tassis, Taylor, \& Wilkinson}]{kiehlmann_compact_2024}
Kiehlmann, S., Lister, M.~L., Readhead, A. C.~S., {et~al.} 2024, The Astrophysical Journal, 961, 240, \dodoi{10.3847/1538-4357/ad0c56}

\bibitem[{{Kokubo} \& {Minezaki}(2020)}]{Kokubo_2020MNRAS.491.4615K}
{Kokubo}, M., \& {Minezaki}, T. 2020, \mnras, 491, 4615, \dodoi{10.1093/mnras/stz3397}

\bibitem[{{Komossa} {et~al.}(2020){Komossa}, {Grupe}, {Gallo}, {Poulos}, {Blue}, {Kara}, {Kriss}, {Longinotti}, {Parker}, \& {Wilkins}}]{Komossa_2020A&A...643L...7K}
{Komossa}, S., {Grupe}, D., {Gallo}, L.~C., {et~al.} 2020, \aap, 643, L7, \dodoi{10.1051/0004-6361/202039098}

\bibitem[{{Laha} {et~al.}(2022){Laha}, {Meyer}, {Roychowdhury}, {Becerra Gonzalez}, {Acosta-Pulido}, {Thapa}, {Ghosh}, {Behar}, {Gallo}, {Kriss}, {Panessa}, {Bianchi}, {La Franca}, {Scepi}, {Begelman}, {Longinotti}, {Lusso}, {Oates}, {Nicholl}, \& {Cenko}}]{Laha_2022ApJ...931....5L}
{Laha}, S., {Meyer}, E., {Roychowdhury}, A., {et~al.} 2022, \apj, 931, 5, \dodoi{10.3847/1538-4357/ac63aa}

\bibitem[{{Laha} {et~al.}(2024){Laha}, {Meyer}, {Roychowdhury}, {Becerra Gonzalez}, {Acosta-Pulido}, {Thapa}, {Ghosh}, {Behar}, {Gallo}, {Kriss}, {Panessa}, {Bianchi}, {La Franca}, {Scepi}, {Begelman}, {Longinotti}, {Lusso}, {Oates}, {Nicholl}, \& {Cenko}}]{laha2024}
---. 2024, ApJ, submitted, 1

\bibitem[{{Li} {et~al.}(2022){Li}, {Ho}, {Ricci}, {Trakhtenbrot}, {Arcavi}, {Kara}, \& {Hiramatsu}}]{li2022}
{Li}, R., {Ho}, L.~C., {Ricci}, C., {et~al.} 2022, \apj, 933, 70, \dodoi{10.3847/1538-4357/ac714a}

\bibitem[{{Longair}(2011)}]{Longair_book}
{Longair}, M.~S. 2011, {High Energy Astrophysics}

\bibitem[{{Masci} {et~al.}(2019){Masci}, {Laher}, {Rusholme}, {Shupe}, {Groom}, {Surace}, {Jackson}, {Monkewitz}, {Beck}, {Flynn}, {Terek}, {Landry}, {Hacopians}, {Desai}, {Howell}, {Brooke}, {Imel}, {Wachter}, {Ye}, {Lin}, {Cenko}, {Cunningham}, {Rebbapragada}, {Bue}, {Miller}, {Mahabal}, {Bellm}, {Patterson}, {Juri{\'c}}, {Golkhou}, {Ofek}, {Walters}, {Graham}, {Kasliwal}, {Dekany}, {Kupfer}, {Burdge}, {Cannella}, {Barlow}, {Van Sistine}, {Giomi}, {Fremling}, {Blagorodnova}, {Levitan}, {Riddle}, {Smith}, {Helou}, {Prince}, \& {Kulkarni}}]{masci2019}
{Masci}, F.~J., {Laher}, R.~R., {Rusholme}, B., {et~al.} 2019, \pasp, 131, 018003, \dodoi{10.1088/1538-3873/aae8ac}

\bibitem[{{Masterson} {et~al.}(2022){Masterson}, {Kara}, {Ricci}, {Garc{\'\i}a}, {Fabian}, {Pinto}, {Kosec}, {Remillard}, {Loewenstein}, {Trakhtenbrot}, \& {Arcavi}}]{masterson2022}
{Masterson}, M., {Kara}, E., {Ricci}, C., {et~al.} 2022, \apj, 934, 35, \dodoi{10.3847/1538-4357/ac76c0}

\bibitem[{{Mathur} {et~al.}(2018){Mathur}, {Denney}, {Gupta}, {Vestergaard}, {De Rosa}, {Krongold}, {Nicastro}, {Collinson}, {Goad}, {Korista}, {Pogge}, \& {Peterson}}]{Mathur_2018ApJ...866..123M}
{Mathur}, S., {Denney}, K.~D., {Gupta}, A., {et~al.} 2018, \apj, 866, 123, \dodoi{10.3847/1538-4357/aadd91}

\bibitem[{Matsumoto \& Piran(2023)}]{matsumoto_generalized_2023}
Matsumoto, T., \& Piran, T. 2023, Monthly Notices of the Royal Astronomical Society, 522, 4565, \dodoi{10.1093/mnras/stad1269}

\bibitem[{Matsumoto \& Piran(2024)}]{matsumoto_late-time_2024}
---. 2024, Late-time {Radio} {Flares} in {Tidal} {Disruption} {Events},  arXiv.
\newblock \url{http://arxiv.org/abs/2404.15966}

\bibitem[{{Meyer} {et~al.}(2018){Meyer}, {Petropoulou}, {Georganopoulos}, {Chiaberge}, {Breiding}, \& {Sparks}}]{meyer2018}
{Meyer}, E.~T., {Petropoulou}, M., {Georganopoulos}, M., {et~al.} 2018, \apj, 860, 9, \dodoi{10.3847/1538-4357/aabf39}

\bibitem[{{Mingo} {et~al.}(2019){Mingo}, {Croston}, {Hardcastle}, {Best}, {Duncan}, {Morganti}, {Rottgering}, {Sabater}, {Shimwell}, {Williams}, {Brienza}, {Gurkan}, {Mahatma}, {Morabito}, {Prandoni}, {Bondi}, {Ineson}, \& {Mooney}}]{mingo2019}
{Mingo}, B., {Croston}, J.~H., {Hardcastle}, M.~J., {et~al.} 2019, \mnras, 488, 2701, \dodoi{10.1093/mnras/stz1901}

\bibitem[{{Nicholls} {et~al.}(2018){Nicholls}, {Brimacombe}, {Kiyota}, {Stone}, {Cruz}, {Trappett}, {Vallely}, {Stanek}, {Kochanek}, {Brown}, {Shields}, {Thompson}, {Shappee}, {Holoien}, {Prieto}, {Bersier}, {Dong}, {Bose}, {Chen}, {Stritzinger}, \& {Holmbo}}]{nicholls2018}
{Nicholls}, B., {Brimacombe}, J., {Kiyota}, S., {et~al.} 2018, The Astronomer's Telegram, 11391, 1

\bibitem[{{Nyland} {et~al.}(2020){Nyland}, {Dong}, {Patil}, {Lacy}, {van Velzen}, {Kimball}, {Sarbadhicary}, {Hallinan}, {Baldassare}, {Clarke}, {Goulding}, {Greene}, {Hughes}, {Kassim}, {Kunert-Bajraszewska}, {Maccarone}, {Mooley}, {Mukherjee}, {Peters}, {Petrov}, {Polisensky}, {Rujopakarn}, {Whittle}, \& {Vaccari}}]{nyland2020}
{Nyland}, K., {Dong}, D.~Z., {Patil}, P., {et~al.} 2020, \apj, 905, 74, \dodoi{10.3847/1538-4357/abc341}

\bibitem[{{O'Dea}(1998)}]{odea1998}
{O'Dea}, C.~P. 1998, \pasp, 110, 493, \dodoi{10.1086/316162}

\bibitem[{{O'Dea} \& {Saikia}(2021)}]{odea2021}
{O'Dea}, C.~P., \& {Saikia}, D.~J. 2021, \aapr, 29, 3, \dodoi{10.1007/s00159-021-00131-w}

\bibitem[{{Panessa} {et~al.}(2019){Panessa}, {Baldi}, {Laor}, {Padovani}, {Behar}, \& {McHardy}}]{panessa2019}
{Panessa}, F., {Baldi}, R.~D., {Laor}, A., {et~al.} 2019, Nature Astronomy, 3, 387, \dodoi{10.1038/s41550-019-0765-4}

\bibitem[{{Perley} \& {Butler}(2017)}]{perley2017}
{Perley}, R.~A., \& {Butler}, B.~J. 2017, \apjs, 230, 7, \dodoi{10.3847/1538-4365/aa6df9}

\bibitem[{{Perlman} {et~al.}(1996){Perlman}, {Stocke}, {Schachter}, {Elvis}, {Ellingson}, {Urry}, {Potter}, {Impey}, \& {Kolchinsky}}]{Perlman_1996ApJS..104..251P}
{Perlman}, E.~S., {Stocke}, J.~T., {Schachter}, J.~F., {et~al.} 1996, \apjs, 104, 251, \dodoi{10.1086/192300}

\bibitem[{{Perrott} {et~al.}(2013){Perrott}, {Scaife}, {Green}, {Davies}, {Franzen}, {Grainge}, {Hobson}, {Hurley-Walker}, {Lasenby}, {Olamaie}, {Pooley}, {Rodr{\'\i}guez-Gonz{\'a}lvez}, {Rumsey}, {Saunders}, {Schammel}, {Scott}, {Shimwell}, {Titterington}, {Waldram}, \& {AMI Consortium}}]{2013MNRAS.429.3330P}
{Perrott}, Y.~C., {Scaife}, A. M.~M., {Green}, D.~A., {et~al.} 2013, \mnras, 429, 3330, \dodoi{10.1093/mnras/sts589}

\bibitem[{{Pushkarev} {et~al.}(2012){Pushkarev}, {Hovatta}, {Kovalev}, {Lister}, {Lobanov}, {Savolainen}, \& {Zensus}}]{pushkarev2012}
{Pushkarev}, A.~B., {Hovatta}, T., {Kovalev}, Y.~Y., {et~al.} 2012, \aap, 545, A113, \dodoi{10.1051/0004-6361/201219173}

\bibitem[{{Raginski} \& {Laor}(2016)}]{raginski2016}
{Raginski}, I., \& {Laor}, A. 2016, \mnras, 459, 2082, \dodoi{10.1093/mnras/stw772}

\bibitem[{{Ramos Almeida} \& {Ricci}(2017)}]{ram2017}
{Ramos Almeida}, C., \& {Ricci}, C. 2017, Nature Astronomy, 1, 679, \dodoi{10.1038/s41550-017-0232-z}

\bibitem[{{Readhead} {et~al.}(1996){Readhead}, {Taylor}, {Xu}, {Pearson}, {Wilkinson}, \& {Polatidis}}]{readhead1996}
{Readhead}, A.~C.~S., {Taylor}, G.~B., {Xu}, W., {et~al.} 1996, \apj, 460, 612, \dodoi{10.1086/176996}

\bibitem[{{Readhead} {et~al.}(1994){Readhead}, {Xu}, {Pearson}, {Wilkinson}, \& {Polatidis}}]{readhead1994}
{Readhead}, A.~C.~S., {Xu}, W., {Pearson}, T.~J., {Wilkinson}, P.~N., \& {Polatidis}, A.~G. 1994, in Compact Extragalactic Radio Sources, ed. J.~A. {Zensus} \& K.~I. {Kellermann}, 17

\bibitem[{{Readhead} {et~al.}(2024){Readhead}, {Ravi}, {Blandford}, {Sullivan}, {Somalwar}, {Begelman}, {Birkinshaw}, {Liodakis}, {Lister}, {Pearson}, {Taylor}, {Wilkinson}, {Globus}, {Kiehlmann}, {Lawrence}, {Murphy}, {O'Neill}, {Pavlidou}, {Sheldahl}, {Siemiginowska}, \& {Tassis}}]{readhead2024}
{Readhead}, A.~C.~S., {Ravi}, V., {Blandford}, R.~D., {et~al.} 2024, \apj, 961, 242, \dodoi{10.3847/1538-4357/ad0c55}

\bibitem[{{Ricci} \& {Trakhtenbrot}(2023)}]{ricci2023}
{Ricci}, C., \& {Trakhtenbrot}, B. 2023, Nature Astronomy, 7, 1282, \dodoi{10.1038/s41550-023-02108-4}

\bibitem[{{Ricci} {et~al.}(2020){Ricci}, {Kara}, {Loewenstein}, {Trakhtenbrot}, {Arcavi}, {Remillard}, {Fabian}, {Gendreau}, {Arzoumanian}, {Li}, {Ho}, {MacLeod}, {Cackett}, {Altamirano}, {Gandhi}, {Kosec}, {Pasham}, {Steiner}, \& {Chan}}]{ricci2020}
{Ricci}, C., {Kara}, E., {Loewenstein}, M., {et~al.} 2020, \apjl, 898, L1, \dodoi{10.3847/2041-8213/ab91a1}

\bibitem[{{Ricci} {et~al.}(2021){Ricci}, {Loewenstein}, {Kara}, {Remillard}, {Trakhtenbrot}, {Arcavi}, {Gendreau}, {Arzoumanian}, {Fabian}, {Li}, {Ho}, {MacLeod}, {Cackett}, {Altamirano}, {Gandhi}, {Kosec}, {Pasham}, {Steiner}, \& {Chan}}]{ricci2021}
{Ricci}, C., {Loewenstein}, M., {Kara}, E., {et~al.} 2021, \apjs, 255, 7, \dodoi{10.3847/1538-4365/abe94b}

\bibitem[{{Roychowdhury}(2023)}]{agniva_thesis}
{Roychowdhury}, A. 2023, PhD thesis, University of Maryland Baltimore County

\bibitem[{{Ryu} {et~al.}(2020){Ryu}, {Krolik}, \& {Piran}}]{ryu2020}
{Ryu}, T., {Krolik}, J., \& {Piran}, T. 2020, \apj, 904, 73, \dodoi{10.3847/1538-4357/abbf4d}

\bibitem[{{Scepi} {et~al.}(2021){Scepi}, {Begelman}, \& {Dexter}}]{scepi2021}
{Scepi}, N., {Begelman}, M.~C., \& {Dexter}, J. 2021, \mnras, 502, L50, \dodoi{10.1093/mnrasl/slab002}

\bibitem[{Sfaradi {et~al.}(2023)Sfaradi, Beniamini, Horesh, Piran, Bright, Rhodes, Williams, Fender, Leung, Murphy, \& Green}]{sfaradi_off-axis_2023}
Sfaradi, I., Beniamini, P., Horesh, A., {et~al.} 2023, Monthly Notices of the Royal Astronomical Society, 527, 7672, \dodoi{10.1093/mnras/stad3717}

\bibitem[{{Shappee} {et~al.}(2014){Shappee}, {Prieto}, {Grupe}, {Kochanek}, {Stanek}, {De Rosa}, {Mathur}, {Zu}, {Peterson}, {Pogge}, {Komossa}, {Im}, {Jencson}, {Holoien}, {Basu}, {Beacom}, {Szczygie{\l}}, {Brimacombe}, {Adams}, {Campillay}, {Choi}, {Contreras}, {Dietrich}, {Dubberley}, {Elphick}, {Foale}, {Giustini}, {Gonzalez}, {Hawkins}, {Howell}, {Hsiao}, {Koss}, {Leighly}, {Morrell}, {Mudd}, {Mullins}, {Nugent}, {Parrent}, {Phillips}, {Pojmanski}, {Rosing}, {Ross}, {Sand}, {Terndrup}, {Valenti}, {Walker}, \& {Yoon}}]{Shappee_2014ApJ...788...48S}
{Shappee}, B.~J., {Prieto}, J.~L., {Grupe}, D., {et~al.} 2014, \apj, 788, 48, \dodoi{10.1088/0004-637X/788/1/48}

\bibitem[{{Shepherd}(1997)}]{1997ASPC..125...77S}
{Shepherd}, M.~C. 1997, in Astronomical Society of the Pacific Conference Series, Vol. 125, Astronomical Data Analysis Software and Systems VI, ed. G.~{Hunt} \& H.~{Payne}, 77

\bibitem[{{Shepherd} {et~al.}(1994){Shepherd}, {Pearson}, \& {Taylor}}]{DIFMAP}
{Shepherd}, M.~C., {Pearson}, T.~J., \& {Taylor}, G.~B. 1994, in Bulletin of the American Astronomical Society, Vol.~26, 987--989

\bibitem[{{Sobolewska} {et~al.}(2023){Sobolewska}, {Siemiginowska}, {Migliori}, {Ostorero}, {Stawarz}, \& {Guainazzi}}]{malgosia2023}
{Sobolewska}, M., {Siemiginowska}, A., {Migliori}, G., {et~al.} 2023, \apj, 948, 81, \dodoi{10.3847/1538-4357/acbb6c}

\bibitem[{Taylor {et~al.}(2000)Taylor, Marr, Pearson, \& Readhead}]{taylor_kinematic_2000}
Taylor, G.~B., Marr, J.~M., Pearson, T.~J., \& Readhead, A. C.~S. 2000, The Astrophysical Journal, 541, 112, \dodoi{10.1086/309428}

\bibitem[{{Tonry} {et~al.}(2018){Tonry}, {Denneau}, {Heinze}, {Stalder}, {Smith}, {Smartt}, {Stubbs}, {Weiland}, \& {Rest}}]{Trony_2018PASP..130f4505T}
{Tonry}, J.~L., {Denneau}, L., {Heinze}, A.~N., {et~al.} 2018, \pasp, 130, 064505, \dodoi{10.1088/1538-3873/aabadf}

\bibitem[{{Trakhtenbrot} {et~al.}(2019){Trakhtenbrot}, {Arcavi}, {MacLeod}, {Ricci}, {Kara}, {Graham}, {Stern}, {Harrison}, {Burke}, {Hiramatsu}, {Hosseinzadeh}, {Howell}, {Smartt}, {Rest}, {Prieto}, {Shappee}, {Holoien}, {Bersier}, {Filippenko}, {Brink}, {Zheng}, {Li}, {Remillard}, \& {Loewenstein}}]{Trakhtenbrot_2019ApJ...883...94T}
{Trakhtenbrot}, B., {Arcavi}, I., {MacLeod}, C.~L., {et~al.} 2019, \apj, 883, 94, \dodoi{10.3847/1538-4357/ab39e4}

\bibitem[{{Tran} {et~al.}(2011){Tran}, {Lyke}, \& {Mader}}]{tran2011}
{Tran}, H.~D., {Lyke}, J.~E., \& {Mader}, J.~A. 2011, \apjl, 726, L21, \dodoi{10.1088/2041-8205/726/2/L21}

\bibitem[{Tremblay {et~al.}(2016)Tremblay, Taylor, Ortiz, Tremblay, Helmboldt, \& Romani}]{tremblay_compact_2016}
Tremblay, S.~E., Taylor, G.~B., Ortiz, A.~A., {et~al.} 2016, Monthly Notices of the Royal Astronomical Society, 459, 820, \dodoi{10.1093/mnras/stw592}

\bibitem[{{van Moorsel} {et~al.}(1996){van Moorsel}, {Kemball}, \& {Greisen}}]{Van_1996ASPC..101...37V}
{van Moorsel}, G., {Kemball}, A., \& {Greisen}, E. 1996, in Astronomical Society of the Pacific Conference Series, Vol. 101, Astronomical Data Analysis Software and Systems V, ed. G.~H. {Jacoby} \& J.~{Barnes}, 37

\bibitem[{{Wang} {et~al.}(2024){Wang}, {Lin}, {Zhang}, \& {Zhu}}]{wang2024}
{Wang}, Y., {Lin}, D. N.~C., {Zhang}, B., \& {Zhu}, Z. 2024, \apjl, 962, L7, \dodoi{10.3847/2041-8213/ad20e5}

\bibitem[{{Wu}(2009)}]{wu2009}
{Wu}, Q. 2009, \mnras, 398, 1905, \dodoi{10.1111/j.1365-2966.2009.15127.x}

\bibitem[{{Zwart} {et~al.}(2008){Zwart}, {Barker}, {Biddulph}, {Bly}, {Boysen}, {Brown}, {Clementson}, {Crofts}, {Culverhouse}, {Czeres}, {Dace}, {Davies}, {D'Alessandro}, {Doherty}, {Duggan}, {Ely}, {Felvus}, {Feroz}, {Flynn}, {Franzen}, {Geisb{\"u}sch}, {G{\'e}nova-Santos}, {Grainge}, {Grainger}, {Hammett}, {Hills}, {Hobson}, {Holler}, {Hurley-Walker}, {Jilley}, {Jones}, {Kaneko}, {Kneissl}, {Lancaster}, {Lasenby}, {Marshall}, {Newton}, {Norris}, {Northrop}, {Odell}, {Petencin}, {Pober}, {Pooley}, {Pospieszalski}, {Quy}, {Rodr{\'\i}guez-Gonz{\'a}lvez}, {Saunders}, {Scaife}, {Schofield}, {Scott}, {Shaw}, {Shimwell}, {Smith}, {Taylor}, {Titterington}, {Veli{\'c}}, {Waldram}, {West}, {Wood}, {Yassin}, \& {AMI Consortium}}]{2008MNRAS.391.1545Z}
{Zwart}, J.~T.~L., {Barker}, R.~W., {Biddulph}, P., {et~al.} 2008, \mnras, 391, 1545, \dodoi{10.1111/j.1365-2966.2008.13953.x}

\end{thebibliography}
\bibliographystyle{aasjournal}

\end{document}